\documentclass[twocolumn, pra, nofootinbib,preprintnumbers,10pt]{revtex4-1}
\usepackage[utf8]{inputenc}
\usepackage{amsmath}
\usepackage{color}
\usepackage{graphicx}
\usepackage{subcaption}
\usepackage{caption}
\usepackage{amssymb}
\usepackage{amsfonts}
\usepackage{hyperref}
\usepackage[format=plain,justification=centerlast,singlelinecheck=true]{caption}

\hypersetup{
colorlinks,
citecolor=blue,
filecolor=blue,
linkcolor=blue,
urlcolor=blue, 
pdfborder = {0 0 0}, 
linktocpage
}

\newcommand{\beq}{\begin{equation}}
\newcommand{\eeq}{\end{equation}}
\newcommand{\beqa}{\begin{eqnarray}}
\newcommand{\eeqa}{\end{eqnarray}}

\def\be{\begin{equation}}

\def\ee{\end{equation}}

\def\eeq{\scriptsize\textmd{eq}}

\def\ttot{\textmd{tot}}

\def\Rre{\textmd{Re}}
\def\Iim{\textmd{Im}}
\def\eeq{\textmd{eq}}

\def\Ttr{\textmd{Tr}}

\def\be{\begin{equation}}
\def\ee{\end{equation}}
\def\bea{\begin{eqnarray}}
\def\eea{\end{eqnarray}}
\def\bit{\begin{itemize}}
\def\eit{\end{itemize}}

\begin{document}

\preprint{LMU-ASC 28/17}


\title{Collective dynamics of accelerated atoms}


\author{Benedikt Richter$^{1,2,3}$}
\author{Hugo Ter\c{c}as$^{1,4}$}
\author{Yasser Omar$^{1,2}$}
\author{In\'es de Vega$^{3}$}

\affiliation{$^1$Instituto de Telecomunica\c{c}\~oes, 1049-001 Lisbon, Portugal}
\affiliation{$^2$Instituto Superior T\'{e}cnico, Universidade de Lisboa, 1049-001 Lisbon, Portugal}
\affiliation{$^3$Arnold Sommerfeld Center for Theoretical Physics, Department f\"ur Physik, Ludwig-Maximilians-Universit\"at, 80333 M\"unchen, Germany}
\affiliation{$^4$Instituto  de  Plasmas  e  Fus\~ao  Nuclear,  1049-001 Lisbon,  Portugal}


\date{November 9, 2017}

\begin{abstract}

We study the collective dynamics of accelerated atoms interacting with a massless field via an Unruh-deWitt type interaction. We first derive a general Hamiltonian describing such a system and then, employing a Markovian master equation, we study the corresponding collective dynamics. In particular, we observe that the emergence of entanglement between two-level atoms  is linked to the building up of coherences between them and to superradiant emission. In addition, we show that the derived Hamiltonian can be experimentally implemented by employing impurities in Bose-Einstein condensates.

\end{abstract}

\maketitle

\section{Introduction}\label{intro}  

The vacuum perceived by a non-inertial observer is by no means a boring place. For instance, it is well-known that an accelerated observer coupled to the vacuum experiences it as a thermal field\footnote{Through out this work, we refer to a quantum field in a thermal (Gibbs) state as a thermal field.} \cite{Unruh:1976db} and that a pair of accelerated atoms starting from a separable state can become entangled \cite{Reznik:2002fz, Brown:2012pw, Salton:2014jaa, Ahmadi:2016fbd}. Furthermore, it is known that for entangled states entanglement can be degraded due to  acceleration, as described for non-localized Fock states \cite{FuentesSchuller:2004xp, Bruschi:2010mc, Richter:2015wha, wang2010quantum} and for localized Gaussian states \cite{Richter:2017noq}. As expected from the equivalence principle, the creation and the degradation of entanglement can also be observed in curved spacetimes \cite{Menezes:2015veo, Dai:2015ota, Kanno:2016gas, Landulfo:2016orf}. Also, acceleration can affect the interactions between two atoms \cite{PhysRevLett.113.020403,PhysRevA.94.012121}. 

In order to analyze these effects, the accelerated atoms are often considered as an open quantum system coupled to a quantum field, which therefore plays the role of an environment \cite{breuerbook, deVega2017}. With this approach, the evolution equation of the atomic reduced density operator -- also known as the master equation -- can be  computed by considering a weak coupling between the atoms and their surrounding field \cite{benatti2004entanglement,  Zhang:2007ha, Hu:2015lda, Moustos:2016lol}. In such master equations, the action of the field in the atomic dynamics is encoded in the dissipative rates, which in turn depend on a sum over environment fluctuations defining the environment correlation function. In this regard, most earlier approaches were based on taking a Wightman correlation function which describe the environment fluctuations as seen from the laboratory frame of reference \cite{Salton:2014jaa,  Hu:2015lda, Zhang:2007ha}.  

In contrast to this, in this paper we consider Rindler spacetime to re-express the Hamiltonian of a collection of uniformly accelerated atoms within the frame of reference of one of them. For the sake of clarity and to avoid technical subtleties, we restrict our analysis to worldlines corresponding to fixed conformal positions of the atoms. This has the advantage that it allows us to explicitly incorporate in the Hamiltonian the red-shifts between atoms having different accelerations. In addition, this representation shows that, for equally accelerated atoms, we recover the Hamiltonian corresponding to standing atoms coupled to a thermal reservoir, as expressed via a thermofield transformation \cite{devega2015}. Hence, our approach provides more physical insight than previous ones, as it allows  comparing the case of accelerated atoms in vacuum with the case of atoms in a thermal field at the Hamiltonian level, i.e., without having to compare the dynamics corresponding to the two cases. Furthermore, the derived Hamiltonian describes an arbitrary number, $N$, of atoms coupled to a massless field via an Unruh-deWitt interaction. This general form enables us to investigate collective effects such as superradiance and to analyze the emergence of entanglement beyond the commonly considered case of only two uniformly accelerated atoms \cite{Reznik:2002fz, benatti2004entanglement, Salton:2014jaa, Hu:2015lda}.

To illustrate our formalism we analyze the dynamics of a collection of $N$ atoms both when they are all equally accelerated and when they have different accelerations. In this context, we explore the conditions for the emergence of cooperative  phenomena and coherent emission. Moreover, considering a Markov and a secular approximation we obtain a Lindblad master equation, which allows us to compute entanglement between the atoms based on a well defined (i.e., positive) reduced density operator. We find that entanglement is indeed built-up during the evolution and persists in the long time limit, an observation that is consistent with previous studies of two uniformly accelerated atoms \cite{benatti2004entanglement}. However, our approach allows us to show that the entanglement itself is not due to the acceleration, which merely produces the effect of a thermal bath in the case of equal accelerations, but rather it is due to the presence of a common environment for the atoms. 

For the accelerations that can be achieved in laboratories, relativistic effects such as the ones described above are generally small \cite{Crispino2008}. Recently, this also was reported in experiments studying photonic entanglement \cite{fink2017experimental}. Therefore, experiments employing analogue systems are more promising candidates to observe these phenomena. Proposed platforms for such experiments include circuit QED \cite{delRey:2011jt, friis2013relativistic}, superconducting qubits \cite{garcia2016entanglement} and cold atoms \cite{PhysRevLett.91.240407, PhysRevD.69.064021, Rodriguez:2016kri}. For the Unruh effect, in particular, an analogue experiment utilizing a Bose-Einstein condensate was proposed \cite{PhysRevLett.91.240407, PhysRevD.69.064021, retzker2008methods}. Recently, also a classical analogue of the Unruh effect was proposed \cite{Leonhardt:2017lwm}. Here, the derived Hamiltonian for the Rindler modes enables us to propose an implementation to simulate the collective dynamics of a collection of co-linearly accelerated atoms based on Bose-Einstein condensates. The key idea is to model the field by the Bogoliubov excitations of the Bose-Einstein condensate and to use optical tweezers to produce artificial two-level atoms. 

The outline of the paper is the following. In Sec.\ \ref{framework}, we introduce the framework that is used in the present work and derive the Hamiltonian describing a system of $N$ accelerated two-level atoms\footnote{In the literature these are sometimes referred to as Unruh-deWitt detectors.} coupled to a scalar field. In Sec.\ \ref{masterequation}, we derive the master equation governing the time evolution of the atoms. In Sec.\ \ref{collectivedyn} we consider such an equation to analyze the atoms' dynamics for different representative cases, and we propose an experimental setup to simulate the open system dynamics in Bose-Einstein condensates. Finally, in Sec.\ \ref{conclusions} we draw the conclusions from this work.

\section{Framework for accelerated atoms}\label{framework}

In this work we consider the setting of several two-level systems, which we refer to as atoms, interacting with a massless scalar field at zero temperature, i.e., in the vacuum state. Assuming that the atoms are initially in the ground state and are in arbitrary inertial motion, it is clear that the system remains in its ground state and no correlations between the atoms can emerge. However, if the atoms are in uniformly accelerated motion, this statement does not remain  true. From the perspective of an observer traveling together with one of the atoms, the field is no longer in the ground state but in an excited state. Therefore, a single atom being accelerated can become excited \cite{Unruh:1976db}, and several atoms interacting with the same field can become correlated \cite{Reznik:2002fz, benatti2004entanglement}. In the following, we study a framework suitable to describe this situation. In particular, we derive the Hamiltonian governing the evolution of many accelerated atoms.

\subsection{Scalar field in Rindler spacetime}

Before moving to the case of Rindler spacetime, we briefly recall some properties of scalar fields in Minkowski spacetime. Let $\phi$ be a massless scalar field confined to a box of length $L$ obeying the Klein-Gordon equation $\Box\phi=0$, where $\Box$ denotes the d'Alembert operator. Then we can expand the field in a complete set of solutions,
\begin{equation}\label{Minkexp} 
\phi=\sum_k \left( a_k u_k(x, t^\text{M})+a_k^\dagger u_k^*(x,t^\text{M})\right),
\end{equation}
where the $u_k(x,t^\text{M})$ are plane-wave solutions of the Klein-Gordon equation that are created and annihilated by the operators $a_k^\dagger$ and $a_k$, respectively. However, in the following, we choose to expand the field $\phi$ in a different complete set of modes that is motivated by the setting we are considering in this work.

\begin{figure}
\begin{subfigure}{\linewidth}
\includegraphics[scale=1.3]{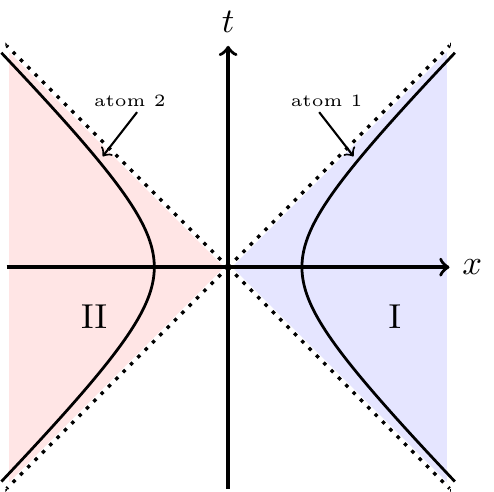}
\caption{Counter-accelerating atoms.} \label{figrindler1}
\end{subfigure}
\vspace{2mm}

\begin{subfigure}{\linewidth}
\includegraphics[scale=1.3]{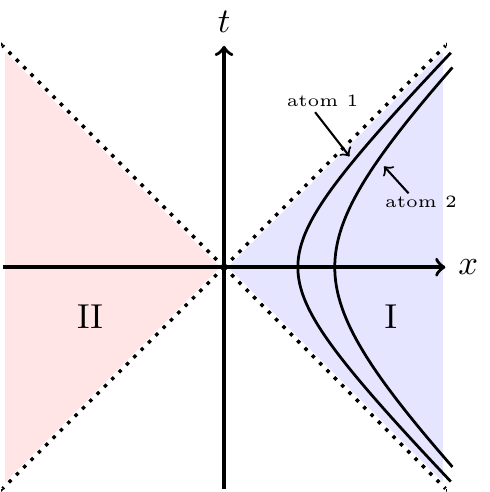}
\caption{Co-accelerating atoms.} \label{figrindler2}
\end{subfigure}
\caption{Schematic figure showing atoms interacting with a massless scalar field. The ground state of the field is entangled between regions $\text{I}$ and $\text{II}$. Therefore the reduced vacuum state in either of the regions is given by a thermal state. In (a), the two atoms are in anti-parallel accelerated motion and interact with the massless field supported in causally disconnected regions of spacetime. In (b), the two atoms are accelerated in parallel and therefore only interact with the field in region $\text{I}$.} \label{figrindler}
\end{figure}

Rindler coordinates are suitable to describe an accelerated observer \cite{Birrell:1982ix, takagi1986vacuum}. In these coordinates a uniformly accelerated object is at rest. Here we discuss the $1+1$ dimensional case, i.e., we neglect the orthogonal Euclidean directions usually labeled by $y$ and $z$. The metric reads in conformal coordinates
\begin{equation}\label{rindlermetric}
ds^2=e^{2\frac{a\xi}{c^2}}\left(c^2 d\tau^2-d\xi^2\right),
\end{equation}
where $\tau$ is the time-like and $\xi$ the space-like coordinate and $a$ is a parameter with the dimension of acceleration. The range of $\tau$  and $\xi$ in each of the wedges is $(-\infty,\infty)$. In Minkowski coordinates the world-line $(x,c t^\text{M})$ of a particle  moving with constant proper acceleration is given by
\begin{subequations}\label{worldlines}
\begin{align}
x=&\pm \frac{c^2}{a} e^{\frac{a\xi}{c^2}}\cosh\left(\frac{a \tau}{c}\right),\\
ct^\text{M}=& \frac{c^2}{a} e^{\frac{a\xi}{c^2}}\sinh\left(\frac{a \tau}{c}\right),
\end{align}
\end{subequations}
where the sign $\pm$ depends on the direction of acceleration; see Fig.\ \ref{figrindler}. The proper acceleration $\alpha$ is related to the acceleration parameter $a$, as $\alpha=a \exp(-a\xi/c^2)$. In consequence, the spatial coordinate $\xi$ is constant and dictated by $\alpha$ for worldlines of uniformly accelerated observers. 

Considering, for instance, two particles with proper accelerations $\alpha_1$ and $\alpha_2$, one obtains the world-lines $(x_1,c t^\text{M}_1)$ and $(x_2,c t^\text{M}_2)$, parametrized by the coordinate time $\tau$ that equals the proper time of particle 1, for $\xi_1=0$, as
\begin{subequations}\label{worldlines2}
\begin{align}
(x_1,c t^\text{M}_1)=& \frac{c^2}{\alpha_1}  \left(\pm\cosh\left(\frac{\alpha_1\tau}{c}\right), \,\sinh\left(\frac{\alpha_1\tau}{c}\right)\right),\\
(x_2,c t^\text{M}_2)=& \frac{c^2}{\alpha_2}  \left(\pm\cosh\left(\frac{\alpha_2\tau}{c}\right),\, \sinh\left(\frac{\alpha_2\tau}{c}\right)\right).
\end{align}
\end{subequations}
To arrive at (\ref{worldlines2}), without loss of generality, we have chosen the acceleration parameter $a$ to coincide with the proper acceleration of particle 1, i.e., $a=\alpha_1$. In consequence, the spacial coordinate of particle 1 is zero, $\xi_1=0$. It may seem that $\alpha_2$ or equivalently the world-line $(x_2,c t^\text{M}_2)$ depend on $\alpha_1=a$. However, that is not the case, as $\xi_2$ is fixed by the proper acceleration of particle 2, $\alpha_2=a \exp(-a\xi_2/c^2)$. Therefore, $\alpha_2$ and $\alpha_1$ are two independent parameters.  What we have done to arrive at (\ref{worldlines2}) is to choose a particular value for the, a priori, unphysical parameter $a$ to obtain a simple form of the world-lines; for an insightful discussion of this issue, we refer to reference \cite{Ahmadi:2016fbd}.

Next, we consider the quantization of a  massless scalar field $\phi$ in this spacetime\footnote{In the following, we work in units where $\hbar=k_B=c=1$.}; see \cite{Birrell:1982ix} for details. A scalar field in a box (size $L$) with periodic boundary conditions can be expanded as
\begin{equation}\label{rindlerexp}
\phi=\sum_k\left( b_k^\text{I} u_k^\text{I}(\xi, \tau)+b_k^{\text{II}} u_k^{\text{II}}(\xi, \tau)\right)+\text{ h.c.}\,,
\end{equation}
where
\begin{subequations}\label{modefunctions}
\begin{align}
u_k^{\text{I}}(\xi, \tau)=& \frac{1}{\sqrt{2L|k|}} e^{i(k\xi-|k|\tau)}, \hspace{5mm} \text{in region \text{I}},\\
u_k^{\text{II}}(\xi, \tau)=& \frac{1}{\sqrt{2L|k|}} e^{i(k\xi+|k|\tau)}, \hspace{5mm} \text{in region \text{II}}
\end{align}
\end{subequations}
are solutions of $\Box\phi=0$, where $\Box$ denotes the d'Alembert operator. The solutions are delta normalized, $(u^\Lambda_k , u^{\Lambda'}_l)=\delta(k-l)\delta_{\Lambda,\Lambda'}$, $(u^\Lambda_k , u^{\Lambda'*}_l)=0$ with $\Lambda,\Lambda'\in\{\text{I},\, \text{II}\}$ \cite{Birrell:1982ix}.\footnote{Here, similar to the quantization of the electromagnetic field, we consider that the field is confined in a box so that the momentum $k$ is quantized. This situation also arises when simulating the system with Bose-Einstein condensates (cf.\ App.\ \ref{siminbecs}), as these are naturally confined to a finite scale $L$ (the size of the condensate). In the continuum limit, corresponding to $L\to\infty$, the normalization of the solutions (\ref{modefunctions}) is modified and the sums over momenta are transformed into integrals.} The time-like Killing vectors in regions $\text{I}$ and $\text{II}$ are given by $\partial_{\tau}$ and $\partial_{-\tau}$ and act on the solutions as $
\partial_{\tau}u_k^\text{I}(\xi, \tau)=  -i |k| u_k^\text{I}(\xi, \tau)$ and $
\partial_{-\tau}u_k^{\text{II}}(\xi, \tau)= -i |k| u_k^{\text{II}}(\xi, \tau)$, i.e., these are the positive frequency solutions. The free field Hamiltonian $H^{\text{I/II}}$ in each of the wedges is given by $H^{\text{I/II}}=\sum_k |k| b_k^{\text{I/II}\dagger} b_k^{\text{I/II}}$. Therefore, the vacuum state in each of the regions $ \text{I/II}$ is  given by $H^{\text{I/II}}|0\rangle_{\text{I/II}}=0$ and the global vacuum can be written as $|0\rangle_\text{R}=|0\rangle_{\text{I}}\otimes|0\rangle_{\text{II}}$. To obtain the complete free field Hamiltonian $H_\text{f}$, one has to take care of the fact that the time-like Killing vectors in regions $\text{I}$ and $\text{II}$, $\partial_{\tau}$ and $\partial_{-\tau}$, differ by a minus sign \cite{Birrell:1982ix}. Therefore, the Hamiltonian also contains a relative minus sign
\begin{equation}\label{freehamiltonian}
H_\text{f}=H^\text{I}-H^{\text{II}}=\sum_k |k|\left( b_k^{\text{I}\dagger} b_k^\text{I}- b_k^{\text{II}\dagger} b_k^{\text{II}}\right).
\end{equation}
Alternatively, the field $\phi$ can also be expanded in a complete set of solutions of the Klein-Gordon equation for Minkowski spacetime. One finds that the Minkowski (M) vacuum $|0\rangle_\text{M}$ is related to the Rindler (R) vacuum by
\begin{equation}\label{minkrindrel}
|0\rangle_\text{M}=S |0\rangle_\text{R},
\end{equation}
where the operator $S$ acts as
\begin{equation}\label{actions}
S^\dagger b_k^{\text{I/II}} S= \cosh(r_k) b_k^{\text{I/II}} +\sinh(r_k) b_k^{\text{II/I} \dagger},
\end{equation}
where $r_k$ is defined by $\tanh(r_k)=e^{-\frac{\pi |k|}{a}}$. In consequence, the Minkowski vacuum expressed in Rindler modes according to (\ref{minkrindrel}) reads
\begin{equation}\label{minkowskivaccum}
|0\rangle_\text{M}=\prod_k\frac{1}{\cosh(r_k)}\sum_{n=0}^\infty \tanh(r_k)^n |n_k\rangle_\text{I} \otimes |n_k\rangle_{\text{II}}.
\end{equation}
Thus, Eq.\ (\ref{minkowskivaccum}) shows that the Minkowski vacuum written in Rindler modes exhibits entanglement between regions $\text{I}$ and $\text{II}$. State (\ref{minkowskivaccum}) is the purification of a thermal state in region $\text{I}$ by modes of region $\text{II}$. Therefore, taking the partial trace over one of the regions results in a thermal state of the Unruh temperature $T_\text{U}\sim \alpha$, where $\alpha$ is the proper acceleration of the observer \cite{Birrell:1982ix}. In consequence, the expectation value of the particle number operator can be expressed, as
\begin{subequations}\label{sinhcosh}
\begin{align}
\sinh(r_{k})^2=&\frac{1}{e^{\beta |k|}-1}\equiv n(k), \label{bedist2}\\
\cosh(r_{k})^2=&\frac{1}{1-e^{-\beta |k|}}=1+n(k),
\end{align}
\end{subequations}
which depends on the inverse Unruh temperature $\beta=\frac{2\pi}{\alpha}$. Written in this suggestive form, (\ref{bedist2}) gives the Bose-Einstein distribution at inverse temperature $\beta$.

\subsection{Hamiltonian for accelerated atoms}

In the following we derive the Hamiltonian governing the evolution of $N$ two-level atoms with rotating frequencies $\omega_i$ coupled to a massless scalar field by an Unruh-deWitt type coupling \cite{Birrell:1982ix}. The interaction is described by an interaction Hamiltonian of the form
\begin{equation}\label{unruhdewitthamiltonian}
H_\text{int}^{(i)}=\chi(\tau_i) Q(\tau_i) \phi[x(\tau_i)],
\end{equation}
where $\chi(\tau_i)$ is the coupling that might be chosen to be constant ($\chi(\tau_i)=\chi=\text{const.}$), $ Q(\tau_i)$ is the monopole moment of the $i$th atom, and $x(\tau_i)$ is its trajectory. The full Hamiltonian for atom $i$, that generates the time translations with respect to the proper time $\tau_i$ of the atom, in the Schr\"odinger picture is given by
\begin{equation}\label{hamforonedetector}
H=\frac{d\tau}{d\tau_i}H_\text{f}+H_{\text{S}}^{(i)}+H_\text{int}^{(i)},
\end{equation}
where $H_{\text{S}}^{(i)}=\omega_i \hat{\sigma}_i^+\hat{\sigma}_i^-$  is the Hamiltonian describing the internal degrees of freedom of the $i$th atom with $\hat{\sigma}^+=\frac{1}{2}(\hat{\sigma}_x +i \hat{\sigma}_y)$ and $\hat{\sigma}^-=\frac{1}{2}(\hat{\sigma}_x-i\hat{\sigma}_y)$, $H_\text{int}^{(i)}$ gives the atom-field interactions, $H_\text{f}$ is the free field Hamiltonian (\ref{freehamiltonian}), $\tau$ is the time coordinate, and $\tau_i$ is the proper time of the atom. A priori, there is no preferred mode expansion for the field $\phi$. The expansion in Minkowski modes, Eq.\ (\ref{Minkexp}), as well as the expansion in Rindler modes, Eq.\ (\ref{rindlerexp}), are both legitimate choices that are equivalent. In this work, contrary to, e.g.\ \cite{Salton:2014jaa}, we choose to formulate the Hamiltonian governing the evolution using the expansion in Rindler modes. We consider the atoms moving along the world-lines $(\xi(\tau_i), \tau(\tau_i))$ introduced in (\ref{worldlines}), where $\xi(\tau_i)=\xi_i$ is fixed by the proper acceleration $\alpha_i$ according to $\alpha_i=a \exp{(-a\xi_i)}$ and the time coordinate reads $\tau=\tau_i\exp{(-a\xi_i)}$. Thus, we identify the red-shift
\begin{equation}
\frac{d\tau}{d\tau_i}=e^{-a\xi_i},
\end{equation} 
and thus, the red-shifted frequencies are defined as 
\begin{equation}
\Omega_i=\frac{d\tau_i}{d\tau}\omega_i.
\label{red-shift} 
\end{equation} 
Even though the chosen family of world-lines is not the most general one\footnote{The set of world-lines of several equally accelerated atoms that at $\tau=0$ are spatially separated, for example, is not contained in the family considered here, since the spacial coordinate $\xi_i$ is fixed by the proper acceleration $\alpha_i$. We also do not consider the case of overlapping Rindler wedges that naturally leads to crossing worldlines. For a discussion of this possibility, see \cite{Ahmadi:2016fbd}.}, in the following, we show that this restricted class of world-lines gives rise to a lot of interesting physical phenomena and that it encompasses various different scenarios involving accelerated atoms. We aim at investigating the general setting of $N$ atoms coupled to a common scalar field $\phi$ according to (\ref{unruhdewitthamiltonian}). Therefore, the Hamiltonian (\ref{hamforonedetector}) has to be generalized to describe more than one atom. It is clear that the free field Hamiltonian $H_\text{f}$ remains unchanged, while the contribution to the energy from the internal dynamics of the atoms is given by the sum of the individual (red-shifted) energies described by $\frac{d\tau_i}{d\tau}H_{\text{S}}^{(i)}$, i.e., the total contribution $H_{\text{S}}$ is given by $H_{\text{S}}=\sum_i \frac{d\tau_i}{d\tau}H_{\text{S}}^{(i)}$. Finally, we have to take care of the individual interaction terms (\ref{unruhdewitthamiltonian}). Also for these, the total interaction energy is given by the sum $H_\text{int}=\sum_i \frac{d\tau_i}{d\tau}H_\text{int}^{(i)}$. Therefore, considering $N$ atoms, we can write the Hamiltonian in the Schr\"odinger picture with respect to the time $\tau$ as
\begin{align}\label{schroedingerham}
&H^{(S)}=\sum_k |k| b_k^{\text{I}\dagger} b_k^\text{I}-\sum_k |k| b_k^{\text{II}\dagger} b_k^{\text{II}}\nonumber\\
+&   \sum_{i=1}^{N_\text{I}} \Omega_i \hat{\sigma}_i^+\hat{\sigma}_i^-- \sum_{i=N_\text{I}+1}^{N} \Omega_i \hat{\sigma}_i^+\hat{\sigma}_i^-\nonumber\\
+&\sum_{i=1}^{N_\text{I}}\sum_k \frac{d\tau_i}{d\tau} \frac{g_{k,i}}{\sqrt{2L|k|}}\left(\hat{\sigma}_i^+ + \hat{\sigma}_i^-\right) \left( b_k^{\text{I}} e^{ik\xi_i}+b_k^{\text{I}\dagger} e^{-ik\xi_i}\right)\nonumber\\
-&\sum_{i=N_\text{I}+1}^{N}\sum_k \frac{d\tau_i}{d\tau}\frac{g_{k,i}}{\sqrt{2L|k|}}\left(\hat{\sigma}_i^+ + \hat{\sigma}_i^-\right) \left( b_k^{\text{II}} e^{ik\xi_i}+b_k^{\text{II}\dagger} e^{-ik\xi_i}\right),
\end{align}
where the first $N_\text{I}$ atoms accelerate in positive direction and the remaining ones accelerate in negative direction, i.e., they live in wedges $\text{I}$ and $\text{II}$, respectively. Further, we defined $g_{k,i}(\tau_i) =\chi(\tau_i) g_{k,i}=\chi g_{k,i}$, where $g_{k,i}$ is the usual coupling appearing in  $Q(\tau_i)$. We note that, in case we consider the Minkowski vacuum, the state of the Rinder modes is given by the entangled state  (\ref{minkowskivaccum}), such that the field in each wedge is in a thermal state, i.e., $\rho_B^{\text{I}}= \Ttr_{\text{II}}\{|0\rangle_{\textmd M}\langle 0|\}\sim \exp{[-\beta\sum_k  |k| b_k^{\text{I}\dagger}b_k^{\text{I}}]}$, and $\rho_B^{\text{II}}=\Ttr_{\text{I}}\{|0\rangle_{\textmd M}\langle 0|\}\sim \exp{[-\beta\sum_k  |k| b_k^{\text{II}\dagger}b_k^{\text{II}}]}$.

Since, throughout this work, we are interested in atoms coupled to a scalar field in the Minkowski vacuum, we frequently encounter vacuum expectation values $\langle H_\text{R} \rangle_\text{M}$ of some Hamiltonian  $H_\text{R}$, defined with respect to Rindler modes. Therefore, it is convenient to absorb the transformation $S$, relating Minkowski and Rindler vacua, into the Hamiltonian and to define transformed Hamiltonians $H_\text{R}'$ by
\begin{equation}\label{vevhamiltonian}
\langle H_\text{R} \rangle_\text{M}=\langle S^\dagger H_\text{R} S \rangle_\text{R}=\langle H_\text{R}'  \rangle_\text{R},
\end{equation}
where the action of $S$ is given in (\ref{actions}). The free Hamiltonian transforms trivially, $H_\text{f}'=H^\text{I}-H^{\text{II}}=H_\text{f}$. The interaction Hamiltonians $H_\text{int}^{(i)}$, however, are not invariant and pick up non-trivial contributions. The transformation of the Hamiltonian $H^{(S)}$ can be performed straight forwardly. The same goes through very similarly for the Hamiltonian in the interaction picture, where  we first transform using the operator $S$ and, subsequently, we go to the interaction picture
\begin{align}
\langle H^{(S)}\rangle_\text{M}=&\langle 0|S e^{i H_0 \tau} H^{(\text{int})'} e^{-i H_0 \tau} S^\dagger|0\rangle_\text{R},
\end{align}
where we introduced the definition
\begin{equation}\label{trafointer2}
H^{(\text{int})'}=e^{-i H_0 \tau} S^\dagger H^{(S)} S e^{i H_0 \tau}.
\end{equation} 
Using expression (\ref{schroedingerham}) for $H^{(S)}$ and definition (\ref{trafointer2}), one obtains for the Hamiltonian in the interaction picture 
{\allowdisplaybreaks
\begin{align}\label{interactionpichamiltonian}
H^{(\text{int})'}=& \sum_{i=1}^{N_\text{I}} \frac{d\tau_i}{d\tau}\bigg[\sum_k  \frac{g_{k,i}}{\sqrt{2L|k|}}\left(\hat{\sigma}_i^+ e^{i \Omega_i\tau} + \text{h.c.}\right) \times\nonumber\\
\times& \bigg(\cosh(r_k)\left( b_k^{\text{I}} e^{i(k\xi_i- |k|\tau)}+b_k^{\text{I}\dagger} e^{-i(k\xi_i- |k|\tau)}\right)\nonumber\\
+&\sinh(r_k)\left( b_k^{\text{II}} e^{i(k\xi_i+|k|\tau)}+b_k^{\text{II}\dagger} e^{-i(k\xi_i+|k|\tau)}\right)\bigg)\bigg]+\nonumber\\
-& \sum_{i=N_\text{I}+1}^{N}\frac{d\tau_i}{d\tau}\bigg[\sum_k \frac{g_{k,i}}{\sqrt{2L|k|}} \left(\hat{\sigma}_i^+ e^{-i \Omega_i\tau} + \text{h.c.} \right)\times\nonumber\\
\times& \bigg(\cosh(r_k)\left( b_k^{\text{II}} e^{i(k\xi_i+|k|\tau)}+b_k^{\text{II}\dagger} e^{-i(k\xi_i+|k|\tau)}\right)\nonumber\\
+&\sinh(r_k)\left( b_k^{\text{I}} e^{i(k\xi_i-|k|\tau)}+b_k^{\text{I}\dagger} e^{-i(k\xi_i-|k|\tau)}\right)\bigg)\bigg].
\end{align}
}
Due to the above transformation, the effective temperature produced by the acceleration is now encoded in the Hamiltonian itself (through the coefficients $\cosh(r_k)$ and $\sinh(r_k)$), while the initial state of the Rindler modes is now the vacuum, $|0\rangle_\text{I}\otimes|0\rangle_{\text{II}}$. Furthermore, the Hamiltonian (\ref{interactionpichamiltonian}) describes an ensemble of $N=N_\text{I}+N_{\text{II}}$ atoms of which $N_I$ are accelerated in one direction and the remaining $N_{\text{II}}$ are accelerated in the opposite direction. Each atom might experience a (different) arbitrary uniform acceleration, where $\xi(\tau_i)=\xi_i$ is constant and fixed by the proper acceleration $\alpha_i$ according to $\alpha_i=a \exp{(-a\xi_i)}$, such that our chosen reference frame moving with atom $1$ sees their frequencies red-shifted according to (\ref{red-shift}). Further, we have considered the definitions (\ref{sinhcosh}) in terms of the inverse Unruh temperature $\beta$.

Our description allows us to read-off that, in the case that all atoms are equally accelerated, the Hamiltonian (\ref{interactionpichamiltonian}) is exactly equivalent to the one describing a collection of $N=N_{\text{I}}$ atoms located at position $\xi_i=\xi>0$ and coupled to a common thermal field in $1+1$ dimensions, once such field is treated with thermofield (also known as thermal Bogoliubov) transformation \cite{israel1976,blasonebook,devega2015}.

Having established the Hamiltonians (\ref{schroedingerham}) and (\ref{interactionpichamiltonian}), we now move on to study the dynamics of the system of accelerated atoms coupled to a massless scalar field. For this purpose, in the next section we derive the respective master equations that govern such evolution in different cases, while in Sec.\ \ref{collectivedyn} we use this equation to numerically analyze the dynamics of up to six accelerated atoms.

\section{Master equation}\label{masterequation}

We derive the master equation obtained by considering a second order perturbative expansion in the coupling Hamiltonian between the atoms and the scalar field. As discussed in Appendix \ref{appenixmastereq}, this equation reads as follows,
\begin{equation}
\frac{d\rho_S (t)}{dt}= - \int_{0}^{t}ds {\textmd{Tr}_B}\{[V_t^0 H^{(\text{int})'} ,[V^0_s H^{(\text{int})'} ,\rho^{\eeq}_{B}\otimes\rho_S (t)]]\},
\label{total6}
\end{equation}
where we refer to the evolution time, given by the proper time $\tau$ of atom 1 as $t$. In this Section, we consider two main situations: all atoms accelerating in the same direction, and then some atoms accelerating in the opposite direction. 

\subsection{Co-accelerating atoms}\label{sameacce}

We insert the Hamiltonian (\ref{interactionpichamiltonian}) with $N_\text{I}=N$ atoms accelerated in one direction and $N_{\text{II}}=0$ atoms accelerated in the opposite direction in the master equation (\ref{total6}). Then, we perform the trace over the environment and consider a change of variables in the time-integrals $t-s\rightarrow s$, such that the resulting equation can be written as
\begin{equation}\label{icc20_NRWA_310}
\frac{d\rho_S (t)}{dt}=
\sum_{i,j}\sum_{\xi,\eta=+,-}\gamma^{\eta\xi}_{ij}(t)[\hat{\sigma}_j^\eta \rho_S(t),\hat{\sigma}_i^\xi]+\textmd{h.c.}
\end{equation} 
with the coefficients $\gamma^{\eta\xi}_{ij}(t)$ defined as 
\begin{equation}
\gamma^{\eta\xi}_{nl}(t)=\int_0^tds C_{nl}(s)e^{\eta i \Omega_l(t-s)} e^{\xi i \Omega_n t} .
\label{coefficients1}
\end{equation}
The correlation functions $C_{nl}(s)$ read
\begin{equation}
C_{nl}(t-s)=\alpha^{\textmd I}_{nl}(t-s)+\alpha^{\textmd{II}}_{nl}(t-s),
\end{equation}
with
\begin{align}
&\alpha^{\textmd I}_{nj}(t-s)=\nonumber\\
&=\sum_k G_{nj}\cosh^2(r_k)\Ttr\{\rho_B^\text{I} b^{\text{I}}_k b^{\text{I}\dagger}_k\}e^{ik(\xi_n-\xi_j)}e^{-i|k|(t-s)}\cr
&=\sum_kG_{nj}\cosh^2(r_k)e^{ik(\xi_n-\xi_j)}e^{-i|k|(t-s)},\nonumber\\
&\alpha^{\textmd{II}}_{nj}(t-s)=\nonumber\\
&=\sum_k G_{nj}\sinh^2(r_k)\Ttr\{\rho_B^\text{II} b^{\text{II}}_k b^{\text{II} \dagger}_k\}e^{ik(\xi_n-\xi_j)}e^{i|k|(t-s)}\cr
&=\sum_k G_{nj}\sinh^2(r_k)e^{ik(\xi_n-\xi_j)}e^{i|k|(t-s)},\label{correl0_12}
\end{align}
where we have defined $G_{nj}=\big(\frac{d\tau_n}{d\tau}\big)\big(\frac{d\tau_j}{d\tau}\big)g_{kn}g_{kj}$, and $\rho_B^{\text{I,II}}=|0\rangle_{\text{I,II}}\langle 0|$ is the vacuum for the modes in I and II, respectively.

We now consider the Markov approximation in the master equation (\ref{icc20_NRWA_310}), which implies that the integral limits of the coefficients (\ref{coefficients1}) are extended to infinity \cite{breuerbook, deVega2017}. As further detailed in Appendix \ref{appenixlongtime}, within this limit the coefficients $\gamma^{\eta\xi}_{jn}(t=\infty)$ can be written as 
\begin{align}
\gamma_{jn}^{+-}=& g_{jn}^{+-}\delta(\Omega_j-\Omega_n)e^{ik_{0j}(\xi_j-\xi_n)},\nonumber\\
\gamma_{jn}^{-+}=& g_{jn}^{-+}\delta(\Omega_j-\Omega_n)e^{ik_{0j}(\xi_j-\xi_n)},
\label{rates_same}
\end{align}
where we have introduced the notation $\gamma_{jn}^{+-}=\gamma_{jn}^{+-}(\infty)$, $\gamma_{jn}^{-+}=\gamma_{jn}^{-+}(\infty)$, with $g_{jn}^{-+}=G_{nj}(n(k_{0j})+1)$, $g_{jn}^{-+}=G_{nj}n(k_{0j})$, and the resonant wave-vector $k_{0j}=\Omega_j$, while the number of excitations in the field is given by the Bose-Einstein distribution (\ref{bedist2}) \footnote{Notice also that unlike in higher dimensions, the rates (\ref{rates_same}) do not decay with the distance $r_{jn}$ between atoms $j$ and $n$. This can naively be understood in analogy with an electric field $E=\nabla \varphi$ ($\varphi$: electric potential).  A consequence of Gauss's law in $d$ spacelike dimensions is that $\nabla E$ scales with the distance $r$ as $\nabla E\sim r_{}^{1-d}$ and, therefore, in one dimension, $E$ is constant in regions with vanishing charge density. We note however, that the factor $1+{\textmd{sgn}}(\Delta_{nj})$ ensures that causality is respected, in the sense that atoms only become connected through the field, so that the rates $\gamma^{\gamma\xi}_{jn}$ are non-zero, once their effective separation becomes time-like.}. In terms of these coefficients, the Markovian master equation can be written, back in the Schr{\"o}dinger picture as 
\begin{align}
\frac{d\rho_S (t)}{dt}=&-i[H_S,\rho_S(t)]+
\sum_{i,j}\gamma^{+-}_{ij}[\hat{\sigma}_j^+ \rho_S(t),\hat{\sigma}_i^-]\nonumber\\
+&\sum_{i,j}\gamma^{-+}_{ij}[\hat{\sigma}_j^- \rho_S(t),\hat{\sigma}_i^+]+\textmd{h.c.}.
\label{MarkovME}
\end{align}
In contrast to the second order master equation (\ref{icc20_NRWA_310}), which does not preserve positivity of $\rho_S(t)$, Eq. (\ref{MarkovME}) is in the Lindblad form and therefore preserves not only the trace and the hermiticity but also the positivity of the reduced density matrix. This is an important property that shall be required if we want to calculate quantities such as entanglement.

\subsection{Counter-accelerating atoms}\label{counteracc}

Following the same steps as in Sec.\ \ref{sameacce}, we find, for atoms accelerating in different directions, that their reduced density matrix obeys the master equation
\begin{align}
\frac{d\rho_S (t)}{dt}=&
\sum_{i,j}\sum_{\xi\neq\eta=+,-}\gamma^{\eta\xi}_{ij}(t)[\hat{\sigma}_j^\eta \rho_S(t),\hat{\sigma}_i^\xi]\nonumber\\
+&\sum_{\kappa,\gamma}\sum_{\xi\neq\eta=+,-}\gamma^{\eta\xi}_{\kappa\gamma}(t)[\hat{\sigma}_\gamma^\eta \rho_S(t),\hat{\sigma}_\kappa^\xi]\nonumber\\
-&\sum_{i,\kappa}\sum_{\xi=\eta=+,-}\gamma^{\eta\xi}_{\kappa i}(t)[\hat{\sigma}_i^\eta \rho_S(t),\hat{\sigma}_\kappa^\xi]\nonumber\\
-&\sum_{i,\kappa}\sum_{\xi=\eta=+,-}\gamma^{\eta\xi}_{i\kappa}(t)[\hat{\sigma}_\kappa^\eta \rho_S(t),\hat{\sigma}_i^\xi]+\textmd{h.c.},
\label{icc20_NRWA_30_first}
\end{align}
where the coefficients $\gamma^{\eta\xi}_{\dots}(t)$ are defined similarly to the ones in the case of parallel acceleration, as detailed in Appendix \ref{appenixcorrelatinfct}. In the long time limit, we find that the equation shall be written as 
\begin{align}
&\frac{d\rho_S (t)}{dt}=
\sum_{i,j}\gamma^{+-}_{ij}(t)[\hat{\sigma}_j^+ \rho_S(t),\hat{\sigma}_i^-]\nonumber\\
+&\sum_{i,j}\gamma^{-+}_{ij}(t)[\hat{\sigma}_j^- \rho_S(t),\hat{\sigma}_i^+]+\sum_{i,j}\gamma^{+-}_{\kappa\gamma}(t)[\hat{\sigma}_\gamma^+ \rho_S(t),\hat{\sigma}_\kappa^-]\nonumber\\
+&\sum_{i,j}\gamma^{-+}_{\kappa\gamma}(t)[\hat{\sigma}_\gamma^- \rho_S(t),\hat{\sigma}_\kappa^+]-\sum_{i,\kappa}\gamma^{+-}_{\kappa i}(t)[\hat{\sigma}_i^+ \rho_S(t),\hat{\sigma}_\kappa^-]\nonumber\\
-&\sum_{i,\kappa}\gamma^{-+}_{\kappa i}(t)[\hat{\sigma}_i^- \rho_S(t),\hat{\sigma}_\kappa^+]-\sum_{i,\kappa}\gamma^{+-}_{i\kappa}(t)[\hat{\sigma}_\kappa^+ \rho_S(t),\hat{\sigma}_i^-]\nonumber\\
-&\sum_{i,\kappa}\gamma^{-+}_{i\kappa}(t)[\hat{\sigma}_\kappa^- \rho_S(t),\hat{\sigma}_i^+]+\textmd{h.c.},
\label{icc20_NRWA_30}
\end{align}
where we have defined $\gamma^{-+}_{ij}(t)$, $\gamma^{+-}_{ij}(t)$, $\gamma^{+-}_{\kappa\gamma}(t)$ and $\gamma^{-+}_{\kappa\gamma}(t)$ as in Eq.\ (\ref{rates_same}), and 
\begin{align}
\gamma_{\kappa i}^{+-}=& \tilde{g}_{\kappa i}^{+-}\delta(\Omega_i-\Omega_\kappa)e^{ik_{0i}(\xi_i-\xi_\kappa)},\nonumber\\
\gamma_{i\kappa}^{-+}=& \tilde{g}_{i\kappa}^{-+}\delta(\Omega_i-\Omega_\kappa)e^{ik_{0i}(\xi_i-\xi_\kappa)},
\label{rates_counter}
\end{align}
with $\tilde{g}^{+-}_{\kappa i}=\tilde{g}^{-+}_{\kappa i}=G_{\kappa i}\sqrt{n(k_{0i})}\sqrt{1+n(k_{0i})}$ and $\tilde{g}^{+-}_{i\kappa}=\tilde{g}^{-+}_{i\kappa}=G_{ i\kappa}\sqrt{n(k_{0i})}\sqrt{1+n(k_{0i})}$.
We note that the cross-rates (\ref{rates_counter}) for counter-accelerating atoms are in general non-vanishing and may give rise to entanglement as described in \cite{Brown:2012pw, Salton:2014jaa}.

\section{Example with six atoms}\label{collectivedyn}

Having developed all the necessary tools, in this section we analyze the collective dynamics of six two-level atoms as viewed from the instantaneous rest frame of the atom $j=1$. We consider for simplicity that all atoms are in the excited state and are accelerated in the same direction, such that $N=N_I$ in the Hamiltonian (\ref{interactionpichamiltonian}), and that the reference atom has a frequency $\omega_1=\omega_s$.  We use units such that $\hbar=k_B=c=1$. Further, time is measured in units of $\omega_s^{-1}$. In consequence, a proper acceleration of $\alpha=1$ corresponds to $\alpha\approx 3\times 10^{21}m/s^2$ for $\omega_s=10^{13}s^{-1}$.

\begin{figure}
 \centering
 \includegraphics[scale=0.45]{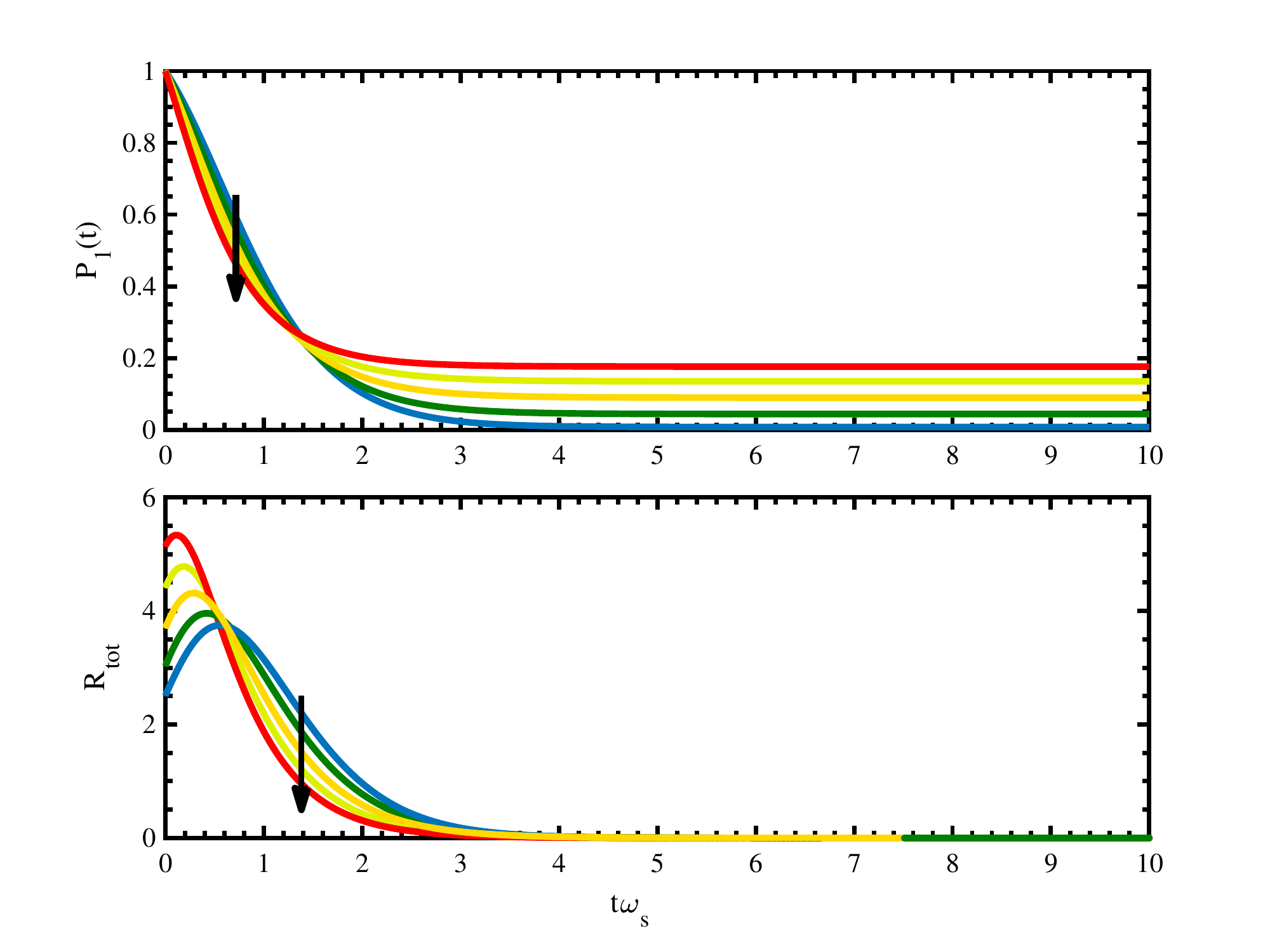}
\caption{Evolution of the atomic population (top panel) and emission rate (bottom panel) for $N=6$ atoms at the same acceleration. The different curves correspond to increasing values of the acceleration $\alpha=2,4,6,8,10$ in a rainbow scale going from blue to red curves (in direction of the arrow). In the top panel, higher accelerations also have higher values of the long time limit population, while in the lower panel, higher accelerations correspond to a higher maximum in the emission rate.}
\label{picGaussian_2}
\end{figure}

\begin{figure}
 \includegraphics[scale=0.45]{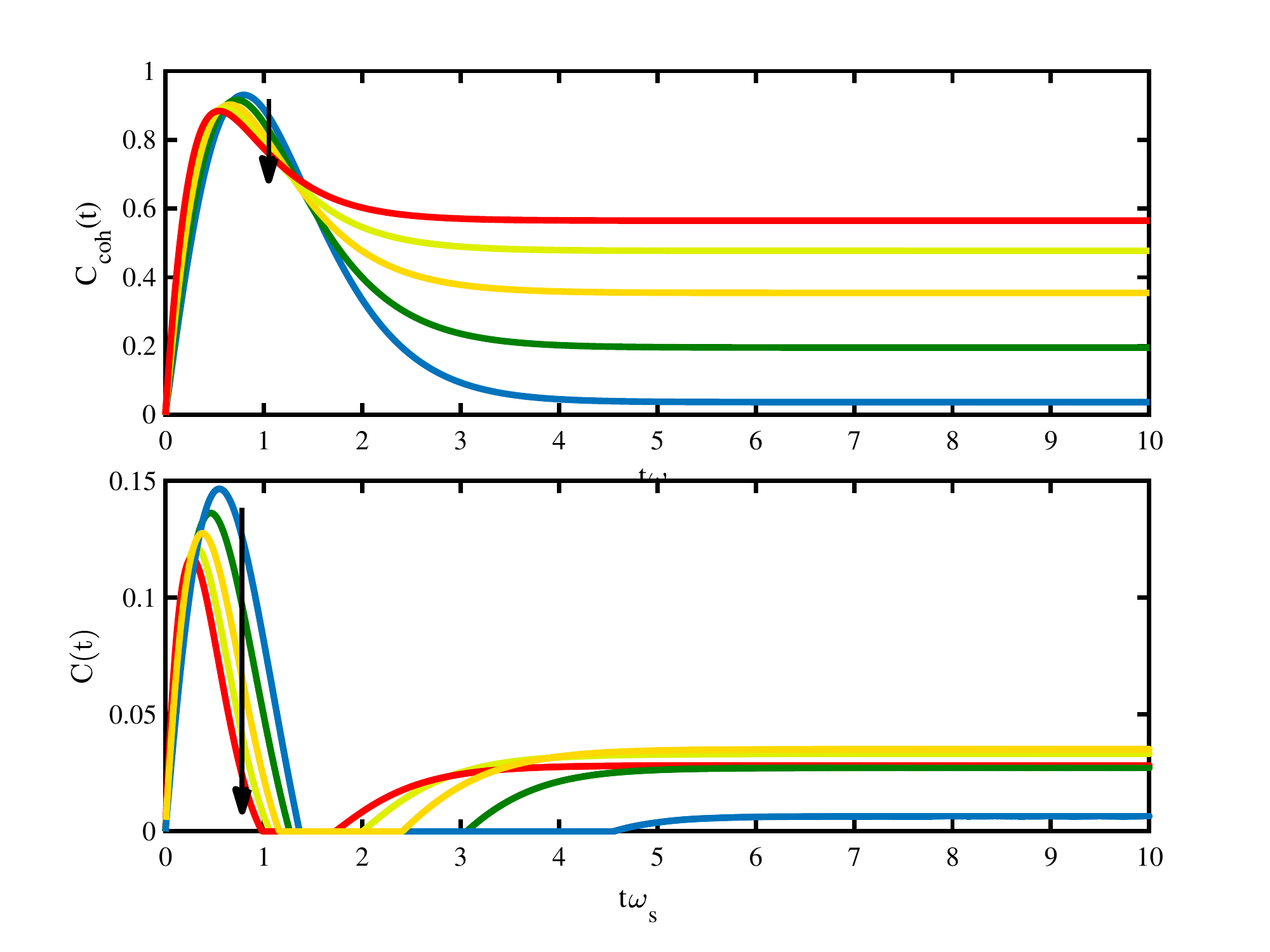}
\caption{Evolution of the atomic coherences (top panel) and concurrence (bottom panel) for $N=6$ atoms at the same acceleration. The different curves correspond to increasing values of the acceleration $\alpha=2,4,6,8,10$ (blue to red curves, in direction of the arrow). In both panels, the curves with higher acceleration have lower maxima than the ones with lower acceleration.}
\label{Figconcurrence}
\end{figure}

In Figs.\ \ref{picGaussian_2} and \ref{Figconcurrence} we consider additionally that atoms have the same acceleration. 
As can be seen in the top panel of Fig.\ \ref{picGaussian_2}, the population of the reference atom $j=1$ 
\begin{equation}
P_1(t)=\langle\Psi_0| \hat{\sigma}_1^+(t)\hat{\sigma}_1^-(t)|\Psi_0\rangle, \label{PC1}
\end{equation} 
evolves to a steady state which contains a finite population in the excited level. Such population is higher the higher the acceleration is. Moreover, as we already noted above, in the case that all atoms experience the same acceleration, their dynamics is equivalent to the dynamics of atoms coupled to a thermal field that is formulated employing a thermofield transformation \cite{israel1976,blasonebook,devega2015}. 

However, in general the steady state achieved is not a thermal state $\rho_S^{th}=e^{-\beta H_S}/Z_S$. Indeed, even if the thermal state is a fixed point of the master equation (\ref{MarkovME}) it is not its only possible steady state due to the presence of symmetries in the system \cite{thingna2016}. This can be confirmed by analyzing the spectrum of the Lindblad superoperator ${\mathcal L}$ corresponding to such an equation (\ref{MarkovME}) when written in the vector form, i.e., $d\rho_S/dt={\mathcal L}\rho_S$. In more detail, when the zero eigenvalue of ${\mathcal L}$ is degenerate, there is no unique steady state. As a consequence, different initial states will evolve into different steady states, and not necessarily to a single and unique (in this case thermal) steady state. In our case of a collection of atoms coupled to a common field, the zero eigenvalue of ${\mathcal L}$ is degenerate, and therefore the system does not always thermalize. We note also that such degeneracy is only removed when the emission rates of the equation are such that $\gamma^{\eta\xi}_{lj}\sim \delta_{lj}\gamma^{\eta\xi}_{ll}$, so that the atoms are virtually coupled to independent environments and evolve independently from the others. 

The fact that atoms are coupled to a common environment also has important consequences for the emission rate 
\begin{equation} \label{emission_rate}
R_{\textmd{tot}}(t)=-\frac{d P_{\textmd{tot}}(t)}{dt}=-\sum^N_{j=1}\frac{d\langle \hat{\sigma}_j^+\hat{\sigma}_j^-\rangle}{dt},
\end{equation}
where $P_{\textmd{tot}}(t)=\sum_jP_j(t)$ and $P_j(t)$ is the population of the atom $j$. As can be observed in the bottom panel of Fig.\ \ref{picGaussian_2}, the atomic emission rate increases for sufficiently small times. This is a clear signature of superradiance \cite{devega2008,navarrete2010}. As can also be observed, such a slope, as well as the location of the superradiant peak (given by the maximum of $R(t)$), is highly dependent on the value of the atomic acceleration. In general, it can be concluded that collective effects are stronger (and therefore the superradiant peak occurs later) the smaller the acceleration is.

The presence of collective effects in the emission is related to the building up of coherences in the atomic system. Such coherences can be quantified in many different ways, such as, for instance, by considering the sum of the off-diagonal elements of the reduced density matrix as proposed in \cite{PhysRevLett.113.140401},
\be
C_{\textmd{coh}}(t)=\sum_{j\neq l}|\langle \hat{\sigma}_j^+\hat{\sigma}_l^-\rangle|.\label{PC}
\ee
In turn, coherences are also related to the generation of entanglement in the atomic ensemble. However, even though coherences are necessary for entanglement to exist, they are not sufficient. That is, a density matrix can have non-vanishing off-diagonal elements (coherences) and still there can be zero entanglement. Therefore, we  also consider here the concurrence for a pair of atoms of the ensemble $j=1,2$. As described in \cite{wootters1998}, the concurrence is defined as 
\begin{equation}
C(\rho_S)=\textmd{max}\{0,\lambda_1-\lambda_2-\lambda_3-\lambda_4\},
\end{equation}
and the $\lambda_i$'s are the eigenvalues, in decreasing order, of the Hermitian matrix $R= \sqrt{\sqrt{\rho_S}\tilde{\rho}_s\sqrt{\rho_S}}$, with $\tilde{\rho}_S=(\sigma^1_y\otimes\sigma^2_y)\rho_S^*(\sigma^1_y\otimes\sigma^2_y)$, where $\rho^*_s$ is the complex conjugated of the reduced density matrix. 

Fig.\ \ref{Figconcurrence} displays the evolution of the coherences (top panel) and the concurrence (bottom panel) for the same value of accelerations as in Fig.\ \ref{picGaussian_2}. In general, coherences are built up in the system around the time at which the emission rates achieve their maximum. However, the persistence of such coherences in the steady state is more significant the higher the acceleration is. In addition, the amount of entanglement encoded in such coherences, as quantified by the concurrence, presents also a growth at initial times of the evolution and shows a higher maximum the smaller the acceleration is. In contrast, at longer times the entanglement appears to be more persistent for accelerations $\alpha\geq 4$.

\begin{figure}
 \includegraphics[scale=0.45]{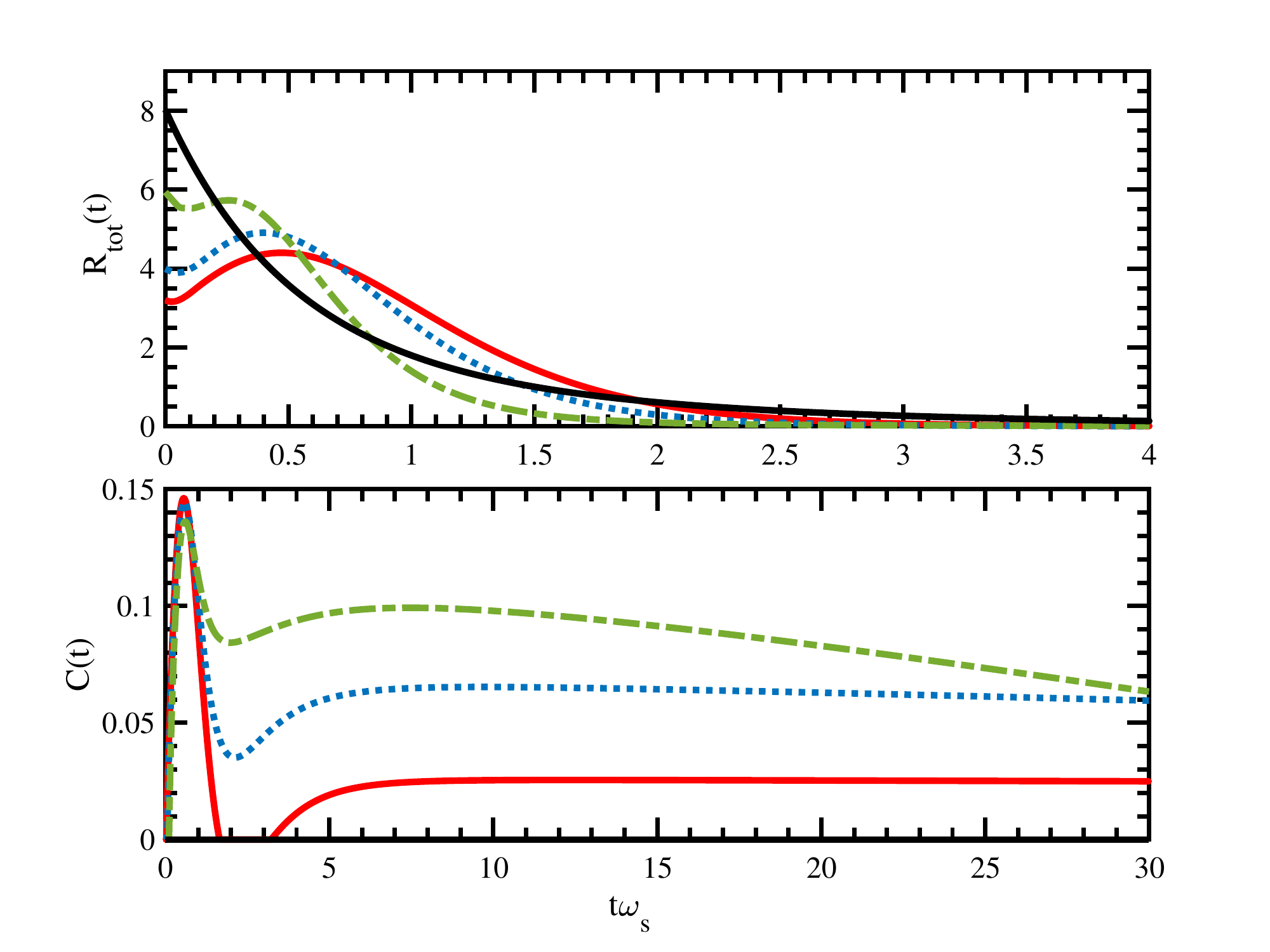}
\caption{Evolution of the emission rate (top panel) and concurrence (bottom panel) for different acceleration distributions, and considering all atoms initially excited. Solid red and solid black curves correspond, respectively, to the cases (a) where all atoms have the same acceleration $\alpha=2$ and (b), where all atoms have different accelerations given by Eq.\ (\ref{alphas}) with $\Delta\alpha=0.6$ and $\omega_j=\omega_1=1$ for all $j$. In the latter case, no entanglement is generated. The dashed green and dotted blue curves corresponds to the case (c) where atoms have different accelerations, as given by Eq.\ (\ref{alphas}), with $\Delta\alpha=0.6$ and $\Delta\alpha=0.03$, respectively. In this case the atomic frequencies are chosen such that $\Omega_j=\omega_1=1$.}
\label{acc_distrib}
\end{figure}

When a set of atoms with equal frequencies experience different accelerations their dynamics can no longer be mapped to that of atoms (having equal frequencies, too) coupled to a common thermal field. Instead, as shown in Fig.\ \ref{acc_distrib}, they present features that are unique to such a system. Focusing in particular on the dynamics of the emission rate and the concurrence, we analyze in these figures the following situations: (a) all atoms having the same acceleration (as considered in the previous Figs.\  \ref{picGaussian_2} and \ref{Figconcurrence}) $\alpha=2$; (b) all atoms $j$ having different accelerations given by
\begin{equation}
\alpha_j=0.2+\Delta\alpha (j-1),
\label{alphas}
\end{equation}
where we have defined $\Delta\alpha$ as an acceleration mismatch parameter. We also consider that all atoms have the same frequency $\omega_j=\omega_1$, such that $\Omega_j\neq \Omega_l$ for any $j\neq l$. In this case, the decaying rates (\ref{rates_same}) are such that $\gamma^{\eta\xi}_{lj}\sim \delta_{lj}\gamma^{\eta\xi}_{ll}$, and each atom evolves independently to the others and relaxes to a thermal state. Finally, in (c) we consider the case where all atoms have different accelerations, but we chose $\omega_j=\exp(-a\xi_j)\,\omega_1$, such that $\Omega_j=\Omega_1=\omega_1$. With such a resonant condition, even when the atoms are accelerated differently, the rates $\gamma^{\eta\xi}_{lj}$ are also non-vanishing for $l\neq j$ and collective effects are still present in the dynamics. However, as shown by Fig.\ \ref{acc_distrib}, superradiance disappears when the acceleration mismatch $\Delta\alpha$ is too large.

For counter-accelerating atoms, coherences are also built between atoms accelerated in opposite directions. However, such coherences do not produce superradiant effects in the emission. To be more specific, when considering an initial state that has no coherences, so that the initial state is separable along the bipartition in atoms in wedge I and atoms in wedge II, both sets of atoms become entangled as time evolves. This is a well-known phenomenon that was studied in a lot of detail \cite{Reznik:2002fz, Salton:2014jaa, Ahmadi:2016fbd, Richter:2017noq, PhysRevA.71.042104}. However, as analyzed numerically (not shown here), the atomic population and the emission rates are not sensitive to the growth of such inter-wedge coherences, and therefore these quantities behave in each region independently of each other. This can also be understood by considering the Heisenberg equations for the correlations
\begin{align}
&\frac{d\langle \hat{\sigma}_l^+ \hat{\sigma}^-_n\rangle}{dt}=\sum_j \gamma_{lj}^{+-} \langle\hat{\sigma}^z_l \hat{\sigma}^-_n \hat{\sigma}_j^+\rangle \nonumber\\
-&\sum_j\gamma_{nj}^{-+} \langle \hat{\sigma}_l^+\hat{\sigma}^z_n \hat{\sigma}^-_j\rangle
+ \sum_j (\gamma_{nj}^{+-})^* \langle \hat{\sigma}^-_j \hat{\sigma}_l^+\hat{\sigma}^z_n\rangle\nonumber\\
 -&\sum_j(\gamma_{lj}^{-+} )^*\langle \hat{\sigma}_j^+\hat{\sigma}^z_l \hat{\sigma}^-_n\rangle
+\sum_\kappa \gamma_{n\kappa}^{-+} \langle\hat{\sigma}^+_l \hat{\sigma}^z_n \hat{\sigma}_\kappa^-\rangle \nonumber\\
-&\sum_\kappa\gamma_{l\kappa}^{+-} \langle \hat{\sigma}_l^z\hat{\sigma}^-_n \hat{\sigma}_\kappa^+\rangle
+\sum_\kappa (\gamma_{l\kappa}^{-+})^* \langle\hat{\sigma}_\kappa^+\hat{\sigma}^z_l \hat{\sigma}^-_n \rangle \nonumber\\
-&\sum_\kappa(\gamma_{n\kappa}^{+-})^* \langle\hat{\sigma}_\kappa^- \hat{\sigma}_l^+\hat{\sigma}^z_n \rangle.
\end{align}
In this regard, for $n\neq l$, the evolution of $\langle \hat{\sigma}_l^+ \hat{\sigma}^-_n\rangle$ depends on correlations between the atoms in different wedges. However, the populations $\langle \sigma^+_n \sigma^-_n\rangle$, do not, since the relevant terms proportional to correlations between the wedges, $\Re\left(\langle\hat{\sigma}^+_l \hat{\sigma}^z_n \hat{\sigma}_\kappa^-\rangle\right)$ and $\Re\left(\langle \hat{\sigma}_l^z\hat{\sigma}^-_n \hat{\sigma}_\kappa^+\rangle\right)$, cancel each other.

Analyzing the physics of an accelerated atomic ensemble is experimentally challenging. This is because for accelerations that can be achieved in laboratories, relativistic effects are typically negligibly small. However, the derived Hamiltonian (\ref{schroedingerham}) can be implemented by considering, for instance, Bose-Einstein condensates (BEC). In Appendix \ref{siminbecs} we give an experimentally feasible scheme to simulate this system that allows us to observe the collective effects displayed in Figs.\ \ref{picGaussian_2} -- \ref{acc_distrib}. 
In particular, the proposed scheme allows us to simulate the case of atoms accelerating in the same direction as described by Hamiltonian (\ref{schroedingerham}). 
To this order, we consider a BEC to play the role of the reservoir (more precisely, the excitations on top of the BEC, the so-called Bogoliubov modes) and a set of impurities immersed in the BEC, which play the role of two-level atoms. The impurities are affected by the potential created by a set of optical tweezers, which provides the ability to tune their internal energies \cite{tweezers}. BECs of alkali atoms are specially suited for the quantum simulation for two reasons. First, the Bogoliubov spectrum in the long-wavelength limit is linear, $\omega\sim  k$ (with $k$ being the wave vector labeling the Bogoliubov mode), which naturally mimics Unruh radiation. Secondly, the kHz energy scale of the BEC excitations is suitable to couple to the two-level atoms created by the optical tweezers. Employing these tweezers a large number of two-level systems can be created on demand having different energy gaps, coupling strengths with the field, and relative spatial positions. Further details of our proposal are given in Appendix \ref{siminbecs}.

\section{Conclusions}\label{conclusions}

In this work we have studied the emergence of collective effects and entanglement in an ensemble of uniformly accelerated two-level atoms. We have derived a Hamiltonian which describes the system in the accelerated frame of reference of one of the atoms. We have shown that, in the limit in which all atoms are equally accelerated, this Hamiltonian is exactly equivalent to the one describing a collection of $N$ atoms coupled to a thermal field once such a thermal field is treated with thermofield or thermal Bogoliubov transformation \cite{israel1976,blasonebook,devega2015}.

 We found that superradiance emerges in an ensemble of accelerated atoms as it is witnessed by the presence of a positive slope in the atomic emission rate at initial times. The slope as well as the location of the subsequent maximum of the emission rate (superradiant peak) varies strongly with the value of the atomic accelerations. In this regard, we found that with higher accelerations the superradiant peak occurs later and therefore collective effects are stronger. As it is also shown, the emergence of such collective effects is linked to the building up of coherences in the atomic system.

Moreover, our formalism shows that the creation of coherences and entanglement within the two-level atoms is due to the fact that they are coupled to a common field, and not directly a consequence of their acceleration. Indeed, for atoms equally accelerated, the acceleration merely converts the surrounding field into an effective thermal field. In this regard, entanglement can be built-up even if the atoms are not accelerating (such that the effective temperature of the field is zero) provided that they are initially in an excited state (see discussion of entanglement generation in common fields in \cite{deVega2017}). Obviously, entanglement can not be created if atoms are initially in their ground state and they have zero acceleration. Thus, acceleration, which leads to a finite temperature field, becomes a fundamental resource to create entanglement only in the case when the initial state of the atoms is the ground state.

However, when atoms having certain frequencies undergo different accelerations the situation is more complex and the dynamics present features that do not correspond to the case of atoms with the same frequencies and coupled to a common thermal field. Thus, the physics of atoms experiencing different accelerations can not be observed in any other scenario than the relativistic one, unless a simulator is specifically designed for this purpose. We have given a concrete  proposal for such a simulator of multiple co-linearly accelerated atoms based on Bose-Einstein condensates. The key idea is to simulate the Unruh radiation field by the Bogoliubov modes (BEC excitations) and to implement the artificial atoms with optical tweezers. Interestingly, the latter setup is not limited to the simulation of collective effects and entanglement generation but also offers the possibility to simulate other effects such as entanglement degradation in accelerated atoms.

\begin{acknowledgements}

We are very grateful to Jad Halimeh and Geza Giedke for enlightening discussions and advice during the development of this work. B.R. and Y.O. acknowledge support from the Physics of Information and Quantum Technologies Group through Funda\c{c}\~{a}o para a Ci\^{e}ncia e a Tecnologia (Portugal), namely through programmes PTDC/POPH/POCH and projects UID/EEA/50008/2013, IT/QuSim, IT/QuNet, ProQuNet, partially funded by EU FEDER, from the EU FP7 project PAPETS (GA 323901), and from the JTF project NQuN (ID 60478). The work of B.R. was supported by the DP-PMI and FCT through scholarship SFRH/BD/52651/2014. H.T. acknowledges the Security of Quantum Information Group (SQIG) for the hospitality and for providing the working conditions, and thanks Fundação para a Ciencia e a Tecnologia (Portugal) through the grants SFRH/BPD/110059/2015 and IF/00433/2015. I.D.V. was financially supported by the Nanosystems Initiative Munich (NIM) (project No. 862050-2) and partially from the Spanish MINECO through project FIS2013-41352-P and DFG-grant GZ: VE 993/1-1.

\end{acknowledgements}

\appendix

 \section{Master equation}\label{appenixmasterequation}

 \subsection{Derivation of the master equation}\label{appenixmastereq}

In this Appendix, we give the details of the derivation of the master equation (\ref{total6}) used in Sec.\ \ref{masterequation}. 

The von Neumann equation for the density operator of the total system in the interaction picture, $\rho^{(\text{int})         }_{tot}(t)$, reads as follows:
\begin{equation}
\frac{d\rho^{I}_{\ttot}(t)}{dt}=\frac{1}{i}[V_t^0 H^{(\text{int})'},\rho^{(\text{int})}_{\ttot}(t)],
\label{total1}
\end{equation}
where we have defined 
\begin{equation}
V_{t-t_0}^0 H^{(\text{int})'} ={\mathcal U}^{-1}_{0}(t,t_0)H^{(\text{int})'} {\mathcal U}_{0}(t,t_0), 
\end{equation}
and also $\rho^{(\text{int})}_{\ttot}={\mathcal U}^{-1}_{0}(t,t_0)\rho(t){\mathcal U}_{0}(t,t_0)$ with the free evolution operator ${\mathcal U}_{0}(t,t_0)=e^{-iH_{0}(t-t_0)}$. To simplify the notation, we set $\rho^{(\text{int})}_{\ttot}(t)=\rho(t)$. We can integrate (\ref{total1}) between $t_0$ and $t$. After two iterations and a trace over the environmental degrees of freedom, this leads to the following equation,
\begin{align}
&\Delta \rho_S (t)=\frac{1}{i}\int_{t_0}^{t}d\bar{\tau} {\textmd{Tr}_B}\{[V^0_{\bar{\tau}} H^{(\text{int})'},\rho(t_0)]\}+ {\left( \frac{1}{i} \right)}^2 \nonumber\\
\times&\int_{t_0}^{t}d{\bar{\tau}}\int_{t_0}^{{\bar{\tau}}} d{\bar{\tau}}' {\textmd{Tr}_B}\{[V^0_{\bar{\tau}} H^{(\text{int})'},[V^0_{{\bar{\tau}}'}H^{(\text{int})'},\rho({\bar{\tau}}')]]\},
\label{total2}
\end{align}
where $\rho_S (t)={\textmd{Tr}_B}\{\rho(t)\}$ is the system reduced density operator and
\begin{equation}
\Delta \rho_S (t)=\rho_S (t)-\rho_S (t_0).
\end{equation}
Eq.\ (\ref{total2}) is exact, but some assumptions have to be made in order to express it as a closed equation for $\rho_S(t)$. For an initially uncorrelated state of the form $\rho(t_0)=\rho_S (t_0)\otimes \rho_B$, and considering the case where
\begin{equation}
\Ttr_B\{V^0_{t_0} H^{(\text{int})'} \rho_B^{\eeq}\}=0,
\label{odd} 
\end{equation} 
the first term in (\ref{total2}) can be eliminated. Note that this occurs for instance when the environment is initially in thermal equilibrium, $\rho_B=\rho_B^{\eeq}=\frac{e^{-\beta H_B}}{\Ttr_B\{e^{-\beta H_B}\}}$.

After the change of variable $T={\bar{\tau}}$ and $s={\bar{\tau}}-{\bar{\tau}}'$, Eq.\ (\ref{total2}) becomes
\begin{eqnarray}
&&\rho_S(t)=\rho_S (t_0)-\int_{t_0}^{t}dT\int_{0}^{T-t_0} ds {\textmd{Tr}_B}\,\{[V^0_{T} H^{(\text{int})'}  ,\cr
&\times&[V^0_{T-s} H^{(\text{int})'}  ,\rho(T-s)]]\}.
\label{total3chap3}
\end{eqnarray}
The evolution equation for the reduced density operator can be obtained by deriving (\ref{total3chap3}) with respect to $t$, 
\begin{equation}
\frac{d\rho_S (t)}{dt}= - \int_{0}^{t-t_0}ds {\textmd{Tr}_B}\bigg\{[V_t^0 H^{(\text{int})'} ,[V^0_{t-s}H^{(\text{int})'}  ,\rho(t-s)]]\bigg\},
\label{total4}
\end{equation}
with initial condition $\rho_S (t_0)$. 
The density operator appearing in the right hand side of (\ref{total4}) has the general form 
\begin{equation}
\rho(t)=\rho_S (t)\otimes \rho_B(t)+\chi_{SB}(t).
\label{totalrho}
\end{equation}
However, the term $\chi_{SB}(t)$, which describes the correlation between the system and the environment at time $t$, can be neglected with the assumption that $\tau_{C}\ll \Delta t$, where $\tau_C$ is the environment correlation time. Such time defines the time that the environment takes to return to its equilibrium state after interacting with the system, and therefore defines also the time scale at which system-environment correlations persist. Neglecting $\chi_{SB}(t)$ corresponds to the Born approximation, which is valid only up to order $g^2$ in the perturbation parameter \cite{atomphotoninteractions,breuer1999,deVega2017}. Also, in order to transform the resulting equation into a time-local form, we further replace $\rho_S(t-s)=\rho_S(t)$ within the integral term. This approximation is valid provided that the system evolution time $T_A$ is much slower than the correlation time of the environment $\tau_C$, which settles the scale in which the integral decays. This is sometimes referred to as the first Markov approximation in the literature. 

Choosing $t_0 =0$, the evolution equation (\ref{total4}) becomes, after a trivial change of variable $s'\rightarrow t-s$,
\begin{equation}
\frac{d\rho_S (t)}{dt}= - \int_{0}^{t}ds' {\textmd{Tr}_B}\{[V_t^0 H^{(\text{int})'} ,[V^0_{s'}H^{(\text{int})'} ,\rho_{B}(t)\otimes\rho_S (t)]]\},
\label{total5}
\end{equation}
where $\rho_B (t)=Tr_{S}\{\rho(t)\}$, and the initial condition is $\rho_S(0)$. This equation can be further simplified by considering that the environment always remains in its equilibrium state, $\rho_B(t)\approx \rho_B^{\eeq}$. Considering this, we find our basic model of equation to consider
\begin{equation}
\frac{d\rho_S (t)}{dt}= - \int_{0}^{t}ds {\textmd{Tr}_B}\{[V_t^0 H^{(\text{int})'},[V^0_sH^{(\text{int})'} ,\rho^{\eeq}_{B}\otimes\rho_S (t)]]\}.
\end{equation}

\begin{widetext}
 \subsection{The dissipative rates for co-accelerating atoms}
 \label{appenixlongtime}

Considering the long time limit of  equation (\ref{icc20_NRWA_310}) implies that the integral limits of the coefficients (\ref{coefficients1}) can be extended to infinity,
\begin{align}
\gamma_{jn}^{-+}(\infty)=&\int_0^\infty ds (\alpha^{\textmd I}_{jn}(t-s)+\alpha^{\textmd{II}}_{jn}(t-s))e^{-i\Omega_js+i\Omega_n t}\nonumber\\
=&\int_0^\infty ds (\alpha^{\textmd I}_{jn}(s)+\alpha^{\textmd{II}}_{jn}(s))e^{i\Omega_js-i(\Omega_j-\Omega_n) t}\nonumber\\
=&\delta(\Omega_j-\Omega_n)\int_0^\infty ds \alpha^{\textmd{I}}_{jn}(s)e^{i\Omega_js},
\end{align}
where we have considered that in the long time limit, $e^{i\Omega_js-i(\Omega_n-\Omega_j) t}$ leads to a non-vanishing contribution only when the phase is zero, i.e., when $\Omega_n=\Omega_j$. Considering the definition in Eq. (\ref{correl0_12}) and going to the continuum limit, we can write
\begin{equation}
\gamma_{jn}^{-+}(\infty)=\delta(\Omega_j-\Omega_n)\int_{-\infty}^{\infty} \frac{dk}{2\pi}\int_0^\infty ds G_{nj}\cosh^2(r_k)e^{ik(\xi_n-\xi_j)}e^{i\Omega_js-i|k|s}.
\end{equation}
Separating now the negative and positive integrals in $k$, we can rewrite the above as 
\begin{equation}
\gamma_{jn}^{-+}(\infty)=\delta(\Omega_j-\Omega_n)\int_{0}^{\infty} \frac{dk}{\pi}2\cos(k(\xi_n-\xi_j))\int_0^\infty ds G_{nj}\cosh^2(r_k)e^{i\Omega_js-i|k|s}.
\end{equation}
Considering now that $\int_0^\infty dt e^{i\omega t}=\pi\delta(\omega)+iP(1/\omega)$, we find
\begin{equation}
\gamma_{jn}^{-+}(\infty)=\delta(\Omega_j-\Omega_n)G_{nj} \bigg[2\cos(k_{0j}(\xi_n-\xi_j))\cosh^2(r_{k_{0j}})(n(k_{0j})+1)+i\frac{2}{\pi }P\int_{-\infty}^\infty dk\bigg(\frac{\Rre\{e^{ik(\xi_j-\xi_n)}J_j(k)(n(k)+1)\}}{|k|-\Omega_j}\bigg)\bigg],
\label{A14}
\end{equation}
where we have defined the resonant wave vector $k_{0j}=\Omega_j$, and extended the limits of the principal value part of the integral to $-\infty$, which can be done given the fact that the integrand is even. Going now to the frequency representation, we rewrite Eq. (\ref{A14}) as
\begin{align}
\gamma_{jn}^{-+}(\infty)=&\delta(\Omega_j-\Omega_n)G_{nj}\times\nonumber\\ 
\times& \bigg[2\cos(k_{0j}(\xi_n-\xi_j))\cosh^2(r_{k_{0j}})(n(k_{0j})+1)+i\frac{2}{\pi}P\int_{-}^\infty d\omega\bigg( \frac{\Rre\{e^{ik(\omega)(\xi_j-\xi_n)}J_j(k(\omega))(n(k(\omega))+1)\}}{\omega-\Omega_j}\bigg)\bigg],
\label{A15}
\end{align}
\end{widetext}
We now consider the Kramers-Kronig relationship,
\bea
\Iim[f(\omega_0)]=-\frac{1}{\pi}P\bigg[\int_{-\infty}^{\infty}d\omega\frac{\Rre[f(\omega)]}{\omega-\omega_0}\bigg],
\eea
which replaced in Eq. (\ref{A15}) leads to the desired result,
\be
\gamma_{jn}^{-+}(\infty)=g^{+-}\delta(\Omega_j-\Omega_n)e^{ik_0(\xi_j-\xi_n)},
\ee
where we have defined the coupling strength $g^{+-}=2G_{nj}(n(k_0)+1)$. In a similar way we find that 
\be
\gamma_{jn}^{+-}(\infty)=g^{-+}\delta(\Omega_j-\Omega_n)e^{ik_0(\xi_j-\xi_n)},
\label{coefficients_same}
\ee
where now $g^{+-}=2G_{nj}n(k_0)$. In addition, we have defined the number of excitations in the field as Eq.\ (\ref{bedist2}). 
Note that similarly as in the quantum optical case, due to causality these rates are non-vanishing only when $t\ge (\xi_j-\xi_n)$. However, such a causality condition is not directly captured by the Markov approach used here (see, for instance, the comment in \cite{PhysRevA.2.883}) but rather should be considered as an ad-hoc condition. This condition is particularly important when the values of $\xi_j-\xi_n$ involved are large compared to the evolution time, which is not the case in our numerical examples. A nice discussion on the retardation effects that exists when connecting different emitters through a common field can be found, for instance, in \cite{PhysRevA.93.023808}.

 \subsection{The dissipative rates for counter-accelerating atoms} 
 \label{appenixcorrelatinfct}

Here, we give the coefficients, $\gamma^{\eta\xi}_{\dots}(t)$, appearing in the master equation counter-accelerating atoms, Eq.\ (\ref{icc20_NRWA_30}). We define the coefficients as 
\begin{align}
&\gamma^{\eta\xi}_{ij}(t)=\int_0^tds C_{ij}(s)e^{\eta i \Omega_j(t-s)} e^{\xi i \Omega_i t},\nonumber\\
&\gamma^{\eta\xi}_{i\kappa}(t)=\int_0^tds C_{i\kappa}(s)e^{-\eta i \Omega_\kappa(t-s)} e^{\xi i \Omega_i t},\nonumber\\
&\gamma^{\eta\xi}_{\kappa i}(t)=\int_0^tds C_{\kappa i}(s)e^{\eta i \Omega_\kappa(t-s)} e^{-\xi i \Omega_i t},\nonumber\\
&\gamma^{\eta\xi}_{\kappa\gamma}(t)=\int_0^tds C_{\kappa\gamma}(s)e^{-\eta i \Omega_\gamma(t-s)} e^{-\xi i \Omega_\kappa t},
\label{coefficients}
\end{align}
where the correlation functions $C_{ij}$ appearing in Eq.\ (\ref{coefficients}) are given by
\begin{equation}
C_{ij}(t-s)=\alpha^{\textmd I}_{ij}(t-s)+\alpha^{\textmd{II}}_{ij}(t-s)
\label{correlation_anti}
\end{equation}
with
\begin{align}
\alpha^{\textmd I}_{ij}(t-s)&=\sum_k G_{ij}\cosh^2(r_k)e^{-i|k|(t-s)},\nonumber\\
\alpha^{\textmd I}_{\kappa\gamma}(t-s)&=\sum_k G_{\kappa\gamma}\sinh^2(r_k)e^{-i|k|(t-s)},\nonumber\\
\alpha^{\textmd I}_{i\kappa}(t-s)&=\sum_kG_{i\kappa}\sinh(r_k)\cosh(r_k)e^{ik(\xi_i-\xi_{\kappa})} e^{-i|k|(t-s)},\nonumber\\
\alpha^{\textmd{II}}_{ij}(t-s)&=\sum_k G_{ij}\sinh^2(r_k)e^{i|k|(t-s)},\nonumber\\
\alpha^{\textmd I}_{\kappa\gamma}(t-s)&=\sum_k G_{\kappa\gamma}\cosh^2(r_k)e^{i|k|(t-s)},\nonumber\\
\alpha^{\textmd{II}}_{i\kappa}(t-s)&=\sum_kG_{i\kappa}\sinh(r_k)\cosh(r_k) e^{ik(\xi_i-\xi_{\kappa})} e^{i|k|(t-s)}.
\label{correl0}
\end{align}
These satisfy the following properties: $\alpha^{\textmd I}_{\kappa i}(t-s)=\alpha^{\textmd I}_{\kappa i}(t-s)=(\alpha^{\textmd I}_{i\kappa}(t-s))^*$ and $\alpha^{\textmd{II}}_{\kappa i}(t-s)=\big(\alpha^{\textmd{II}}_{i\kappa}(t-s)\big)^*$.

\section{Simulation in Bose-Einstein condensates}\label{siminbecs}

In this appendix, focusing on the case of atoms that are accelerated in the same direction, we propose a simulation of the system described in Sec.\ \ref{collectivedyn}. The simulation is based on Hamiltonian (\ref{schroedingerham}).

\subsection{Bogoliubov excitations as Unruh radiation}
We start by implementing the first term of Eq. \eqref{schroedingerham}. A bosonic field $\phi(x,t)$ in a quasi-one-dimensional Bose-Einstein condensate (BEC) is described by the following Hamiltonian 
\begin{equation}
H_f=\int dx~ \phi^\dagger \left[-\frac{\hbar^2}{2m}\frac{\partial^2}{\partial x^2}-\mu+u_0\phi^\dagger\phi\right]\phi, 
\label{eq_ham1}
\end{equation}
where $\mu$ is the chemical potential, $u_0$ is the interaction strength and $a_k$ is a bosonic operator satisfying the usual commutation relation $[a_k, a_{k'}^\dagger]=\delta_{k,k'}$ \cite{pethick2002bose}. In the following, we make use of the Bogoliubov approximation, which amounts to neglecting the depletion from the macroscopically occupied vacuum state $\langle \phi\rangle= \sqrt{n_0}$, such that
\begin{equation}
\phi(x,t)= e^{-i\mu t/\hbar}\left[\sqrt{n_0}+\frac{1}{\sqrt{L}}\sum_k \left(a_k e^{i k x}+a_k^\dagger e^{-ik x}\right)\right].
\label{eq_bog1}
\end{equation}
Plugging this into Eq. \eqref{eq_ham1}, we obtain
\begin{equation}
\begin{array}{l}
\displaystyle{H_f=E_0+\sum_k \epsilon_k a_k^\dagger a_k}\\

\displaystyle{+\frac{1}{2}u_0n_0\sum_{k\neq0}\left(2a_k^\dagger a_k +a_{k}^\dagger a_{-k}^\dagger +a_k a_{-k}\right)},
\end{array}
\label{eq_bog2}
\end{equation}
where $E_0=N u_0 n_0$ is a spurious energy shift, with $N$ denoting the total number of particles in the BEC, and $\epsilon_{k}=\hbar^2k^2/(2 m)$. By performing a Bogoliubov-Valatin transformation of the form $a_k=u_kb_k+v_k^*b_{-k}^\dagger$, the condition of $b_k$ being also a bosonic operator implies the normalization condition $\vert u_k\vert^2-\vert v_k\vert^2=1$, and diagonalizes the Bogoliubov Hamiltonian \eqref{eq_bog2} as
\begin{equation}
H_f=\sum_{k}E_kb_k^\dagger b_k,
\label{eq_bog3}
\end{equation}
where $E_k=\sqrt{\epsilon_k \left(\epsilon_k+2\mu\right)}$ is the energy spectrum, and the Bogoliubov coefficients are given by \cite{pethick2002bose}
\begin{equation}
u_k, v_k=\left(\frac{\epsilon_k+\mu}{2E_k}\pm \frac{1}{2}\right)^{1/2}.
\end{equation}  
The condensate depletion (both quantum and thermal) per mode $k$ is given by $\langle a_k^\dagger a_k\rangle=\vert v_k\vert^2+(\vert u_k\vert^2+\vert v_k\vert^2) n_k$, where 
\begin{equation}
n_k=\frac{1}{e^{\epsilon_k/k_BT}-1}
\end{equation}
is the Bose-Einstein statistics and $T$ is the temperature of the system. Since our goal is to mimic Unruh radiation, we identify the Bogoliubov modes with the Unruh modes propagating in region I of Rindler spacetime with acceleration $\alpha=2\pi c k_B T/\hbar$. As such, we garantee that the first term appearing in the Hamiltonian \eqref{schroedingerham} is accurately simulated by Eq. \eqref{eq_bog3}.

\subsection{Optical tweezers as tunnable two-level systems }

\begin{figure}[t]
\includegraphics[scale=0.42]{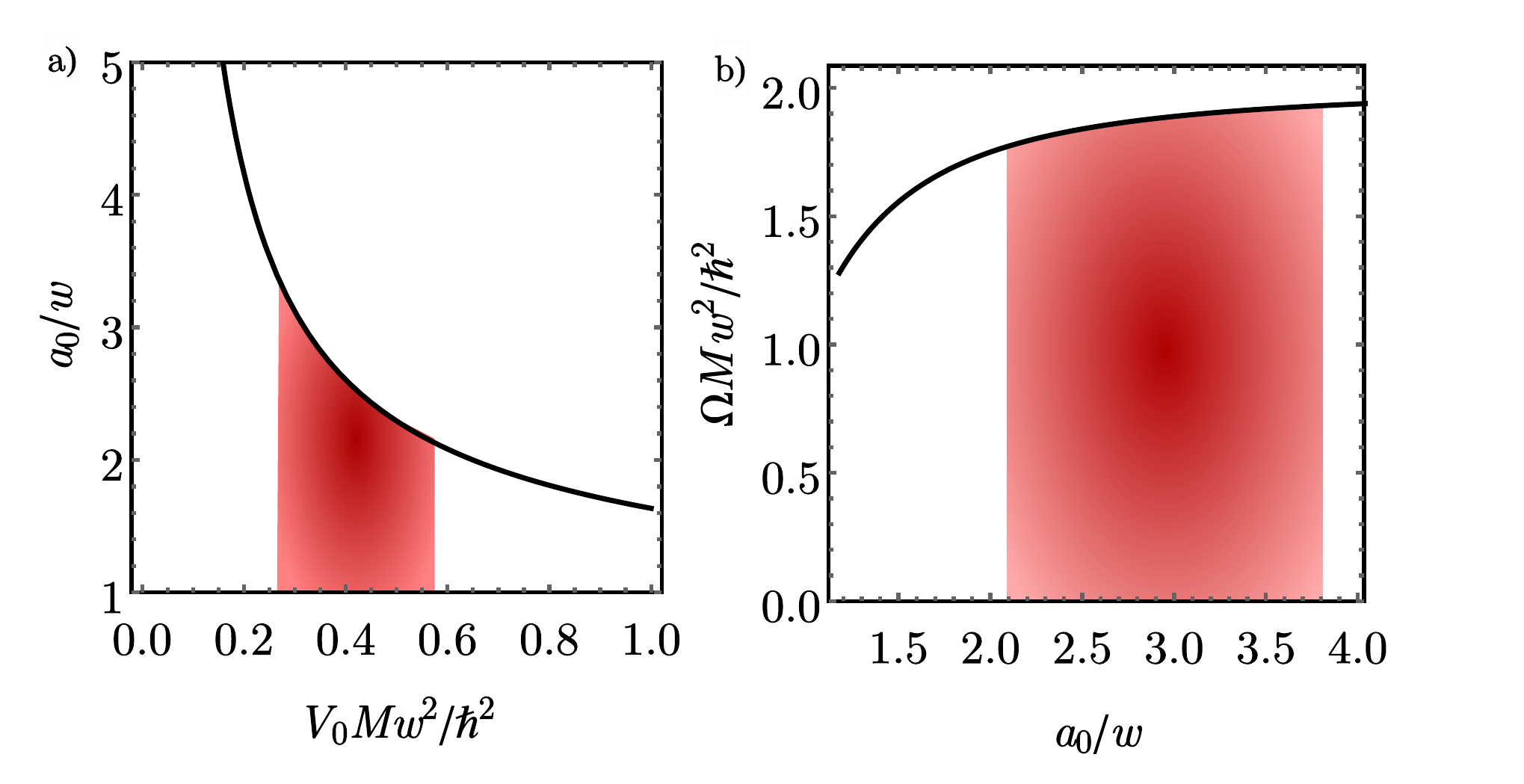}
\caption{a) Width of the bound states as a function of the optical tweezer potential depth. b) Dependence of the atom energy on the bound state width. In both panels, the shadowed region corresponds to the two-level condition in Eq.\ \eqref{eq_twolevel}.}\label{figure5}
\label{figurex}
\end{figure}

In order to emulate the Hamiltonian of the two-level systems (atoms), we make use of a set of optical tweezers, which can be located at different positions on demand. As we are about to see, the spatial distribution of the optical tweezers will simulate the location of the atoms at different positions in Rindler spacetime \cite{kim2016situ}. Let $\psi(x)$ denote the field of an auxiliary particle (impurity) inside the BEC. The corresponding Hamiltonian reads
\begin{equation}
H_S=\int dx ~ \psi^\dagger \left[-\frac{\hbar^2}{2 M}\frac{\partial^2}{\partial x^2}-V_0 e^{-x^2/w^2}\right]\psi,
\label{eq_qubit1}
\end{equation}
where $V_0$ represents the depth of the potential (associated with the laser intensity) and $w$ is the potential width (i.e., the laser beam waist). By expanding the field in the form $\psi(x)=\sum_n \varphi_n(x) c_n$, where $\varphi_n(x)$ satisfies the Schr\"odinger equation $H_S \varphi_n(x)=\varepsilon_n \varphi_n(x)$ and $c_n$ is a bosonic operator with algebra $\left[c_n, c_m^\dagger\right]=\delta_{n,m}$, we can evoke the WKB approximation in order to determine the number of bound states $n_{b}$ as
\begin{equation}
n_{b}=\lfloor \frac{2}{\hbar^2}\sqrt{\frac{V_0 M}{\pi w}}-\frac{1}{2}\rfloor,
\end{equation}
with $\lfloor \cdot\rfloor$ denoting the integer part. We are mostly interested in the case where exactly two bound states can be produced ($n_b=2$); see Fig.\ \ref{figure5}. By keeping the potential depth constant and tuning the tweezer waist, for example, we obtain two-level atoms with energies  in the range
\begin{equation}
\frac{4}{5\hbar^2}\sqrt{\frac{M V_0}{\pi}}<w<\frac{4}{3\hbar^2}\sqrt{\frac{M V_0}{\pi}}.
\label{eq_twolevel}
\end{equation}
In that case, the two bound states $n=0$ and $n=1$ can be approximately described by the following variational wave functions,
\begin{equation}
\varphi_{0}(x)=\left(\frac{2}{\pi a_0^2}\right)^{1/4}e^{-x^2/a_0^2}, \quad \varphi_1(x)=2\frac{x}{a_0}\varphi_0(x),
\end{equation}
where $a_0$ is the width of the bound state, which can be related to the tweezer parameters as
\begin{equation}
\frac{w^2}{2a_0^2}\left(\frac{2}{a_0^2}+\frac{1}{w^2}\right)^3=\frac{V_0^2 M^2 w^4}{\hbar^4}.
\label{eq_tls_condition}
\end{equation}
The variational energies are then given by $\varepsilon_n=\langle \varphi_n\vert H_S \vert \varphi_n \rangle$, and the two-level (atom) transition energy $\Omega=\varepsilon_1-\varepsilon_0$ is then given by
\begin{equation}
\displaystyle{
\Omega=\frac{2\hbar^2}{M a_0^2}-\sqrt{2}V_0\frac{\sqrt{2 a_0^4+\frac{a_0^6}{w^2}}}{\left(a_0^2 +2w^2\right)^2}}.
\end{equation}
Multiple atoms can therefore be simulated by tuning the width $w_i$ of the different tweezers independently, which will then emulate the atoms' Hamiltonian in Eq.\ \eqref{schroedingerham} provided the identification $\omega_i d\tau_i/d \tau \rightarrow \Omega_i$, yielding 
\begin{equation}
H_S=\sum_i\Omega_i \sigma^+_i\sigma^-_i,
\end{equation}
where $\sigma^+_i=c_{i,1}^\dagger c_{i,0}$ and $\sigma^-=c_{i,0}^\dagger c_{i,1}$.

\subsection{System-bath interaction in the rotating-wave approximation}

The interaction between a collection of two-level systems, created by the impurities trapped in the optical tweezers, and the phonons in the BEC is described by the following Hamiltonian
\begin{equation}
H_\text{int}=g\sum_i\int dx~\psi^\dagger(x-x_i)\phi^\dagger(x)\phi(x) \psi(x-x_i),
\end{equation}
where $x_i$ is the location (in the laboratorial frame) of each optical tweezer and $g$ is the atom-atom interaction strength ($g=u_0$ if the atoms and the reservoir are of the same species). By using the expansion in Eq.\ \eqref{eq_bog2} and the two-level condition in \eqref{eq_twolevel}, we obtain 
\begin{equation}
H_\text{int}=H_\text{int}^{(0)}+H_\text{int}^{(1)}+H_\text{int}^{(2)},
\label{eq_int1}
\end{equation}
where $H_\text{int}^{(0)}=g n_0/L\sum_i\sum_{n=0}^{1} c_{i,n}^\dagger c_{i,n}$ is the BEC Stark shift, which can be incorporated by a renormalizing of the energy levels in the form $\varepsilon_n\rightarrow \varepsilon_n+g n_0$. The last term is second-order in the bosonic operators $b_k$, $\mathcal{O}(b_k^2)$, which we neglect in the spirit of the Bogoliubov approximation. Finally, the first-order term can be easily given as
\begin{equation}
H_\text{int}^{(1)}=\sum_k \sum_{j}\sum_{m, n=0}^1\mathcal{G}_{ik}^{mn}e^{ik x_i}c_{i,m}^\dagger b_k c_{i,n} + {\rm h.c.},
\label{eq_int2}
\end{equation}
\begin{figure}[]
\includegraphics[scale=0.46]{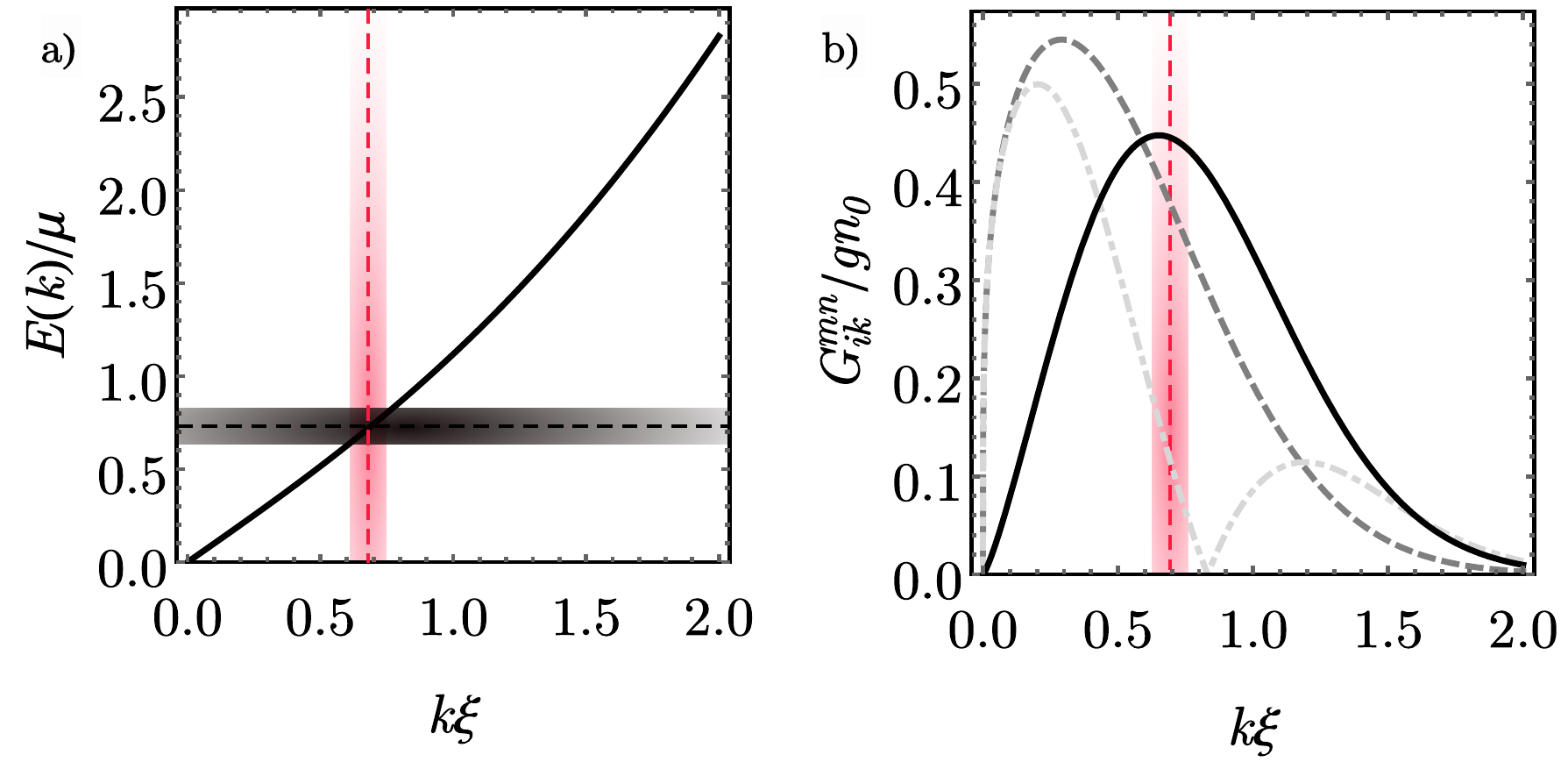}
\caption{a) Dispersion relation of the Bogoliubov modes and the choice of the resonance value of the atoms. The horizontal (vertical) dashed line indicates the frequency (wave number) resonant with the central atom transition $\Omega$. The shadowed horizontal (vertical) stripes indicate the range of near-resonant frequencies (wave vectors) for a centered distribution of atom proper accelerations. b) Strength of the couplings $\vert\mathcal{G}^{00}_{ik}\vert$ (dashed line), $\vert\mathcal{G}^{11}_{ik}\vert$ (dot-dashed line) and $\vert\mathcal{G}^{10}_{ik}\vert$ (solid line) near resonance. For illustration, we have used $a_0=2.2 w$.}
\label{fig_RWA}
\end{figure}
where the components of the coupling tensor explicitly read
\begin{equation}
\begin{array}{c}
\mathcal{G}^{00}_{ik}=g\sqrt{\frac{n_0 S(k)}{L}}e^{-k^2a_0^2/2},\\\\
\mathcal{G}^{11}_{ik}=\left(1-\frac{a_0^2k^2}{2} \right)\mathcal{G}^{00}_{ik},\quad 
\mathcal{G}^{10}_{ik}=\mathcal{G}^{01^*}_{ik}=ia_0 k \mathcal{G}^{00}_{ik}.\\
\end{array}
\end{equation}
Here, $S(k)=u_k-v_k$ denotes the BEC static structure factor within the Bogoliubov approximation. Eq.\ \eqref{eq_int2} contains intra-band ($m=n$) and inter-band ($m\neq n$) terms. However, intra-band couplings involve long wavelength phonons $k\sim 0$, for which $S(k)\sim 0$, and can therefore be neglected \cite{1367-2630-9-11-411}. Moreover, we choose a narrow range of atom energies $\Omega_i$ for which the quasi-resonant $k_i$ modes are located around the maximum of $\vert \mathcal{G}^{01}_{ik}\vert$ (see Fig.\ \ref{fig_RWA} for illustration). We then go to the interaction picture, as described above, to drop the terms proportional to $\sigma_i^+ b_k^\dagger$ and $\sigma_i^- b_k$. Within the RWA approximation, the interaction Hamiltonian finally reads
\begin{equation}
H_{I}^{\rm RWA}\simeq \sum_{i,k} C_{ik}e^{ikx_i}\sigma_i^+ b_k +{\rm h. c.}, 
\end{equation}
where $C_{ik}=\mathcal{G}_{ik}^{10}$. The latter is valid if the coupling between the optical tweezers and the BEC is sufficiently weak, i.e., provided the condition $g\ll u_0$. This is why a different species is necessary, allowing for $g$ to be tuned via Feshbach resonances. The appropriate simulation of the system-bath reservoir in Eq.\ (\ref{schroedingerham}) is performed if we identify the laboratory positions $x_i$ with the Rindler coordinates $\xi_i$ and the matrix element $C_{ik}$ with the term $g_{ik} d\tau_i/d\tau$.  \par

Typical experimental setups with laser powers of $\sim 800$ mW result in potential depths of $V_0\sim 2\pi \times 1$ kHz and beam waist of $w\sim 1.0$ $\mu$m \cite{OptTweezersColdAtoms}. The latter are comparable to the typical values of healing length $\xi_{\text{heal}}$ and chemical potential $\mu$ in elongated $^{87}$Rb condensates \cite{Denschlag97}.  Moreover, the atoms could be constructed with $^{172}$Yb atoms, which are heavy $(M/m\simeq 2)$ and weakly-interacting enough (an estimate of the scattering lengths $a_{\rm Rb-Yb}\sim -160.7 a_{\rm Bohr}$ and $a_{\rm Rb-Rb}\sim 90.0 a_{\rm Bohr}$ yields $g\sim 0.18 u_0$ \cite{PhysRevA.88.052708, PhysRevA.59.1303}) such that the approximations above hold  (see Ref. \cite{PhysRevLett.113.237203} and references therein). In typical $^{87}$Rb experiments with $n_0 \sim 50$ $\mu$m$^{-1}$ (i.e., $\sim5000$ atoms confined in a trap of size 100 $\mu$m \cite{PhysRevLett.105.265302}), and using the reasoning of Ref. \cite{PhysRevLett.113.237203}, we estimate that retardation effects can be neglected for up to $N\sim 20$ atoms separated by $d~\sim 2.8$ $\mu m$. In this case, hopping between the different tweezers can also be prevented.

\end{document}